\documentclass[10pt,aps,prd,twocolumn,showpacs,showkeys]{revtex4}

\usepackage{amsmath,amssymb}
 \usepackage{epsfig}
\usepackage{graphicx}
\usepackage{amsmath}
\usepackage{mathrsfs}
\usepackage[bookmarks=false]{hyperref}
\usepackage[utf8]{inputenc}
\newcommand{\beq}{\begin{equation}}
\newcommand{\eeq}{\end{equation}}
\newcommand{\beqa}{\begin{eqnarray}}
\newcommand{\eeqa}{\end{eqnarray}}
\newcommand{\beqar}{\begin{eqnarray*}}
\newcommand{\eeqar}{\end{eqnarray*}}

\begin{document}

\title{Plethora of relativistic charged spheres: The full spectrum of     
Guilfoyle's static, electrically charged spherical solutions}

\author{Jos\'e P. S. Lemos}
\affiliation {Centro Multidisciplinar de Astrof\'{\i}sica -- CENTRA,
Departamento de F\'{\i}sica,  Instituto Superior T\'ecnico - IST,
Universidade de Lisboa - UL, Avenida Rovisco Pais 1, 1049-001 Lisboa, Portugal
 \\ Email: joselemos@ist.utl.pt}

\author {Vilson T. Zanchin}
\affiliation {Centro de Ci\^encias Naturais e Humanas, Universidade
Federal do ABC, Avenida dos Estados, 5001, 09210-580 Santo Andr\'e, S\~ao 
Paulo,
Brazil \\ Email: zanchin@ufabc.edu.br}

\begin{abstract}

We show that Guilfoyle's exact solutions of the Einstein-Maxwell equations for 
spherical symmetric static electrically charged matter with a 
Reissner-Nordstr\"om exterior possess a bewildering plethora of different 
types of solutions.  For the parameter space of the solutions we use two 
normalized variables, $q^2/R^2$ and $r_0/R$, where $q$ is the total electric 
charge, $r_0$ is the radius of the object, and $R$ is a length representing 
the square root of the inverse energy density of the matter. The two other 
parameters, the mass $m$ and the Guilfoyle parameter $a$, both dependent on 
$q$, $r_0$ and $R$, are analyzed in detail. The full parameter space of 
solutions $q^2/R^2\times r_0/R$ is explored with the corresponding types of 
solutions being identified and analyzed. The different types of solutions are 
regular charged stars, including charged dust stars and stars saturating the 
Buchdahl-Andr\'easson bound, quasiblack holes, regular charged black holes 
with a de Sitter core, regular black holes with a core of phantom charged 
matter, other exotic regular black holes, Schwarzschild stars, Schwarzschild 
black holes, Kasner spacetimes, pointlike and planar naked singularities, and 
the Minkowski spacetime. Allowing for $q^2<0$, in which case it is not 
possible to interpret $q$ as electric charge, also yields new solutions, some 
of which are interesting and regular, others are singular. Some of these types 
of solutions as well as the matter properties have been previously found and 
studied, here the full spectrum being presented in a unified manner.


\end{abstract}


\maketitle


\date{today}
\section {Introduction}
\label{sec-introd}

Guilfoyle found an amazing class of electric spherically symmetric
interior fluid solutions joined to a Reissner-Nordstr\"om outer
spacetime \cite{guilfoyle}. The considerable potential of these starlike
solutions has been understood by Lemos and Zanchin, who studied
the specific aspects of its fluid properties \cite{lemoszanchin2009}, analyzed
quasiblack holes as the frozen stars \cite{lemoszanchin2010}, tested the
sequence of configurations that saturate the Buchdahl-Andr\'easson bound
for gravitational collapse \cite{lemoszanchin2015}, and found regular black 
holes within the whole set of solutions \cite{lemoszanchin2016}.
From these previous works it is clear that Guilfoyle's solutions
possess a richness of solutions that must be explored in full.
We thus embark on a full study of the properties of the bewildering variety
of this class of Guilfoyle's star solutions, i.e., we give the full
spectrum of the solutions.

The motivation for the Guilfoyle's solutions \cite{guilfoyle} starts with the 
work of Weyl \cite{weyl1} who has considered static Einstein-Maxwell systems. 
Static Einstein-Maxwell systems have played an important role in understanding 
the structure of extremely compacted objects in general relativity. The 
relative simplicity of the resulting system of differential equations allied 
to the rich analytical properties resulting from the coupling of the 
electrostatic Maxwell and the gravitational fields are a reason for such a 
relevance. In  Weyl \cite{weyl1} a functional relation between the metric 
component $g_{tt}\equiv B$ and the electric potential $\phi$, $B=B(\phi)$, the 
Weyl ansatz, is assumed to exist. In electrovacuum he found that if there is a 
relation, it must be of a quadratic form $B(\phi)= \left(-\epsilon\phi + 
b\right)^2 +c$, the Weyl relation, where $\epsilon=\pm 1$, $b$ and $c$ are 
arbitrary constants, and we use units such that the speed of light and the 
Newton's gravitational constant equal unity. Majumdar \cite{majumdar} showed 
that such a quadratic relation exists for generic static spacetimes, not only 
for the axisymetric ones that Weyl had used \cite{weyl1}. When $c=0$ the 
relation between the potentials is a perfect square and in that case Majumdar 
\cite{majumdar} and Papapetrou \cite{papapetrou} showed that the solution is 
the vacuum solution exterior to some static charged dust distribution in which 
the gravitational attraction balances exactly the electric repulsion. When the 
 Majumdar-Papapetrou relation holds, i.e., $B(\phi)= \left(-\epsilon\phi + 
b\right)^2$, and besides electric fields there is matter one can further show  
\cite{majumdar,papapetrou} that the pressure of the matter content should 
vanish and that the charge density $\rho_{\rm e}$ and the energy density 
$\rho_{\rm m}$ of the charged dust must obey $\rho_{\rm e}= \epsilon\rho_{\rm 
m}$, i.e., the matter is extremal matter (see also \cite{lemoszanchin2005}). 
One can then construct interior Majumdar-Papapetrou matter solutions matched 
to an exterior Majumdar-Papapetrou vacuum solution. For spherical symmetry 
these solutions are the Bonnor stars \cite{bonnorwickra2,lemoszanchin2008}, 
with an exterior extremal Reissner-Nordstr\"om spacetime. On the other hand, 
if one wants to include fluid pressure into the solutions, the 
Majumdar-Papapetrou relation cannot hold and the Weyl relation runs into 
difficulties. So a new route, other than Weyl's and Majumdar and Papapetrou's, 
was taken. The idea was to generalize the interior spherical symmetric 
Schwarzschild solution, a very interesting and useful solution, and this was 
done by Cooperstock and de la Cruz \cite{cooperstock} and Florides 
\cite{florides}. They found electric star solutions. Guilfoyle final 
motivation was to use $B=B(\phi)$, the Weyl ansatz, and see which functional 
forms would work that would give solutions with pressure \cite{guilfoyle}. A 
simple one is $B(\phi)= a\left(-\epsilon\phi + b\right)^2$, with $a$ a new 
parameter, the Guilfoyle parameter, in which case the Cooperstock and de la 
Cruz \cite{cooperstock} star is reproduced as a particular solution, but there 
are others also used in \cite{guilfoyle}.

The motivation for the work \cite{lemoszanchin2009} is related to the study of 
the fluid properties that obey a Weyl ansatz $B=B(\phi)$, namely, fluids that 
obey either a Weyl relation, a Majumdar-Papapetrou relation, or a Guilfoyle 
relation. The work \cite{lemoszanchin2009} is based, besides \cite{guilfoyle}, 
in the works of Das \cite{das62}, De and Raychaudhuri \cite{deraychaudhuri}, 
Gautreau and Hoffman \cite{gautreau}, and Bonnor \cite{bonnornewton}. Das 
\cite{das62}, showed that if the ratio $\rho_{\rm e}/\rho_{\rm m} = \epsilon$ 
is assumed, then the relation between potentials must be of the form of the 
Majumdar-Papapetrou relation $B =(-\epsilon\phi+ b)^2$. De and Raychaudhuri 
\cite{deraychaudhuri} went a step further and generalized this theorem 
assuming more general conditions, such as there is a closed equipotential 
within the charged dust fluid with no singularities, holes, or another kind of 
matter, then the charged dust fluid corresponds to a Majumdar-Papapetrou 
solution. The inclusion of matter with pressure into the Weyl relation was 
first considered by Gautreau and Hoffman \cite{gautreau}. They verified that 
if the metric potential $B$ is given by the Weyl relation $B(\phi)= 
\left(-\epsilon\phi + b\right)^2 +c$, then the perfect fluid satisfies the 
condition $\rho_{\rm e}\left(\epsilon\phi-b\right)  = -\epsilon 
\left(\rho_{\rm m}+ 3\,p \right)\sqrt B$. Bonnor \cite{bonnornewton} also 
displayed some theorems for these type of systems. In \cite{lemoszanchin2009} 
the Gautreau and Hoffman \cite{gautreau} result was generalized to systems 
obeying a Guilfoyle relation, namely, $a\rho_{\rm 
e}\left(\epsilon\phi-b\right)  = -\epsilon \left(\rho_{\rm m}+ 3\,p + \epsilon 
(1-a)\rho_{\rm em}\right)\sqrt B$, where $\rho_{\rm em}$ is the 
electromagnetic energy density, and some new results drawn upon 
\cite{guilfoyle} were extended.

The motivation for \cite{lemoszanchin2010} was the search for quasiblack holes 
within Guilfoyle's stars \cite{guilfoyle}. Quasiblack holes are stars on the 
verge of becoming black holes, stars whose boundary to the exterior is the 
event horizon. Quasiblack holes have been reported  in 
\cite{lemosweinberg,kleber,lemoszanchin2006}. Their properties were studied in 
\cite{lemoszaslavskii1,lz2,lz3,lz4,lz5}. Previous results with electric stars 
with quasiblack hole properties were reported by Bonnor \cite{bonnorwickra1} 
(see also \cite{bonnorwickra2}) and in \cite{felice1,felice2,bonnor99}, in 
\cite{felice2} a stability analysis of the solutions was performed. Quasiblack 
holes with non-Abelian gauge fields were found in \cite{lw} and quasiblack 
holes with rotation have also been found in \cite{meinel}. Now, in 
\cite{guilfoyle} some simple cases of solutions were displayed,  notably 
solutions which obeyed the condition $m> \frac{q^2}{r_0}$, where $m$, $q$, and 
$r_0$ are the mass, electric charge and radius of the star, respectively. By 
extending these solutions to the limiting case $m= \frac{q^2}{r_0}$ quasiblack 
holes  with pressure, i.e., relativistic charged spheres as frozen stars, were 
found within Guilfoyle's solutions \cite{lemoszanchin2010}.

The motivation for \cite{lemoszanchin2015} was to search whether Guilfoyle's 
stars  \cite{guilfoyle} saturate the Buchdahl-Andr\'easson bound. Buchdahl 
\cite{buchdahl} showed that, under some physical assumptions, a star made 
of a perfect fluid would obey a compactness bound $r_0/m \geq 9/4$. The  
interior Schwarzschild solution obeys this bound, the equality appearing when 
the star's central pressure goes to infinity. For electrically charged matter 
it was found by Andr\'easson \cite{andreasson-charged} that there was also a 
bound $r_0/m \geq 9/\left(1+\sqrt{1+3q^2/r_0^2}\right)^2$. While Buchdahl's 
proof is appropriate for stars, Andr\'easson's proof is appropriate for thin 
shells, and he demonstrated that indeed thin shells saturate his bound, the 
Buchdahl-Andr\'easson bound. In \cite{lemoszanchin2015} it was shown that 
Guilfoyle's stars also saturate the Buchdahl-Andr\'easson bound. There 
are other electric stars that do not saturate the bound, although
they almost do 
\cite{alz,llqz}.

The motivation for \cite{lemoszanchin2016} was to search for regular black 
holes within Guilfoyle's solutions \cite{guilfoyle}. Regular black holes were 
envisioned by Bardeen \cite{bardeen1968} and subsequent works displayed other 
type of solutions and their properties 
\cite{dy92,ab00,br06,mat08,lemoszanchin2011,balaklemoszaya}. In 
\cite{lemoszanchin2016} it was shown that indeed there are also regular black 
holes in Guilfoyle's solutions \cite{guilfoyle}.

Here we perform a full analysis and make a classification of the plethora of 
Guilfoyle's solutions for the relation $B(\phi)= a\left(-\epsilon\phi + 
b\right)^2$, this set was named Ia in \cite{guilfoyle}. We explore the whole 
parameter space and try to interpret all the corresponding solutions. In 
passing we will deal with imaginary electric charges, a concept that appeared 
first in the context of particle solutions in general relativity and is linked 
to the Einstein-Rosen bridge \cite{einsteinrosen}, see also \cite{tidal}, and 
we will also deal with Kasner metrics that arise in the interior of black 
holes \cite{hiscock:1997jt,gpbook}.

The paper is organized as follows. In Sec.~\ref{sec-sphericalGuilfoyle0} the 
exact Guilfoyle's set of solutions with the ansatz $B(\phi)= 
a\left(-\epsilon\phi + b\right)^2$ is displayed in full, with its interior, 
exterior, and junction surface. We also display the constraints the parameters 
should satisfy. In Sec.~\ref{sec-analysis} a general analysis of the mass $m$ 
and the Guilfoyle parameter $a$ of the solutions in terms of the other free 
parameters is done. In Sec.~\ref{sec-regions} we  show that it is useful to 
visualize the parameter space regions in a plot $r_0/R\times q^2/R^2$. The 
regions, i.e., areas in this parameter space, then appear naturally according 
to the characteristics of the respective solutions and we classify them and 
name the corresponding objects, such as normal stars, tension stars, regular 
black holes, singular black holes and others. A brief description of the kinds 
of solutions contained in each region is also given. In 
Sec.~\ref{sec-boundaries} we study the solutions belonging to the boundaries 
of the regions, i.e., areas, described in the previous section. These regions 
are lines and points in the parameter space on which  interesting solutions 
can be  found. Finally, in Sec.~\ref{sec-conclusion} we conclude.

\section{Guilfoyle's solutions and constraints on the solutions}
\label{sec-sphericalGuilfoyle0}

\subsection{Spherical static Weyl-Guilfoyle systems and equations}
\label{sec-sphericalGuilfoyle}

The spacetime is static and spherically symmetric with the metric
written in  the form
\beq
ds^2 = -B(r)\,dt^2 +A(r)\,dr^2+r^2\,(d\theta^2+\sin^2\theta\,d\varphi^2)\, ,
\label{metricsph}
\eeq
where $(t,\,r,\, \theta,\, \varphi)$ are spherical symmetric spacetime 
coordinates, and the functions $A$ and $B$ depend only on the radial 
coordinate $r$.

The source is a static charged fluid distribution with spherical symmetry.
The fluid has an energy density $\rho_{\rm m}(r)$, an isotropic
pressure $p(r)$, an electric charge density $\rho_{\rm e}(r)$, and
a time component of the four-velocity given by
\beq
U_t =  -\sqrt{B(r)\,}\,, \label{velocity}
\eeq
the other components of the four-velocity being zero. The electromagnetic 
gauge potential has only one nonzero component, given by
\beq
 {\cal A}_t = -\phi(r)\,,
\label{electpotential}
\eeq 
with $\phi(r)$ being the electric potential.

The mass $M(r)$ inside a sphere of radius $r$ is defined as
\beq
M(r) = \int_0 ^r 4\pi\, r^2 \left( \rho_{\rm m}(r)
+\frac{Q^2(r)}{8\pi\,r^4}\right)dr + \frac{Q^2(r)}{2\,r} \,,
\label{massdef}
\eeq
and the electric charge $Q(r)$inside a sphere of radius $r$ is defined as
\beq
Q(r) =  4\pi\int_0^r { \rho_{\rm e}(r)\sqrt{A(r)} \,r^2 dr}
\,.
\label{chargedef}
\eeq
These two quantities, $M(r)$ and $Q(r)$, are auxiliary and important 
quantities.

Now, we impose a Weyl ansatz into the system, i.e., the metric potential 
$B(r)$ and the electric potential $\phi(r)$ are functionally related through 
some relation $B=B(\phi)$, or in detail $B(r)=B(\phi(r))$. Here we want to 
study systems that obey the Guilfoyle relation \cite{guilfoyle} (see also 
\cite{lemoszanchin2009}), namely, $B(r)= a 
\left[-\epsilon\,\phi(r)+b\right]^2$, where $\epsilon=\pm 1$, and $a$ and $b$ 
being arbitrary constants. Without loss of generality one can set $b=0$, since 
it can be absorbed into the electric potential $\phi$. Thus, the Guilfoyle 
relation can be written as
\begin{equation}
B(r)= a \phi^2(r) \, , \label{wg-relation}
\end{equation}
with $a$ being called the Guilfoyle parameter.

The Einstein-Maxwell equations provide the set of equations for this system.
The gravitational part of the equations yields the relations,
\beqa
&&  \frac{B^\prime(r)}{B(r)}+\frac{A^\prime(r)}{A(r)} =
8\pi r\,A(r)
\Big[\rho_{\rm m}(r) +p(r)\Big]\, ,\label{einsteq1}\\
&&
 \Big(\frac{r}{A(r)}\Big)^\prime = 1- 8\pi\, r^2\left(
\rho_{\rm m}(r) + \frac{Q^2(r) }{8\pi r^4}\right)\, , \label{einsteq2}
\eeqa
with a prime denoting the derivative with respect to the radial coordinate $r$ 
(units in which the gravitational constant and the speed of light are set to 
one are used). The definition of the total charge $Q(r)$ inside $r$, see 
Eq.~(\ref{chargedef}), gives that the only nontrivial component of the Maxwell 
equation is $Q(r) = {r^{2}\,\phi^\prime (r)}/{\sqrt{B(r) \,A(r)}}$, where an 
integration constant has been put to zero. The electric potential  $\phi (r)$ 
can be written in terms of $B(r)$ as $\epsilon \phi(r) = \sqrt{{B(r)}/{a}}$, 
see Eq.~\eqref{wg-relation}. Then, Eq.~\eqref{chargedef}, can be written as
\beq
Q(r) = \frac{-\epsilon\,r^{2}\,B^\prime(r)} {\,2\sqrt{a A(r)}\, B(r)}\, .
\label{qspherical2}
\eeq
Once one has the metric functions $B(r)$ and $A(r)$ one obtains the electric 
charge distribution through Eq.~\eqref{qspherical2}. 
Equations~\eqref{einsteq1}-\eqref{qspherical2} provide the equations for 
Weyl-Guilfoyle systems.

\subsection{Guilfoyle's solutions}
\label{sec-solution}

\subsubsection{Interior solution}

We assume that the interior solution extends from $r=0$ up to $r=r_0$.
Guilfoyle's solutions are found under the additional assumption that the
effective energy density $\rho_{\rm m}(r) + {Q^2(r)}/{8\pi\,r^4}$
is a constant, i.e., 
\beq
8\pi\,\rho_{\rm m}(r) + \dfrac{Q^2(r)}{r^4} = \frac{3}{R^2}\, ,
\label{rhoconst}
\eeq
where $R$, is a length parameter associated to the inverse of the total 
energy density. Through the junction conditions of the metric at the surface 
$r_0$, one finds that $R$ is to be related to the parameters of the exterior 
solution, namely, the total mass $m$ and the total charge $q$. The 
equation of state \eqref{rhoconst} is a generalization of the interior 
Schwarzschild solution equation of state to include electrically charged 
matter and had been proposed previously \cite{cooperstock,florides}. With this 
additional assumption provided by \eqref{rhoconst}, one can find that 
Guilfoyle's solutions \cite{guilfoyle} are given by 
\beq
A(r) =\left({1 - \dfrac{r^2}{R^2}}\right)^{-1}\, , \label{A-sol1}
\eeq
\beq
B(r)  = \left[\frac{\left(2-a\right)^2}{a^2}\,F^2(r)
\right]^{a/(a-2)} \, ,
\label{B-sol1}
\eeq
\beq
\phi(r) = \epsilon \sqrt{\dfrac{B(r)}{a}}\,,
\label{phi00}
\eeq
\beq
8\pi \rho_{\rm m}(r) = \frac{3}{R^2} -\frac{a}{\left(2-a\right)^2}
\frac{k_0^2}{R^4}\,\frac{r^2} {F^2(r)}\, , 
\label{rhom-sol1} 
\eeq
\beqa
8\pi p(r) &=& -\frac{1}{R^2} + \frac{a}{\left(2-a\right)^2}
\frac{k_0^2}{R^4}\,\frac{r^2} {F^2(r)}      +\nonumber\\ &&
+ \frac{2a}{2-a}\dfrac{k_0}{R^2}\frac{\sqrt{1-\frac{r^2}{R^2}}}{F(r)}\, ,
\label{p-sol1}
\eeqa
\beq
4\pi \rho_{\rm e}(r) = 
\frac{\epsilon \sqrt{a}}{2-a}\frac{k_0^2}{R^4}\frac{r^2}{F^2(r)}\!
\left(\!1+
\frac{3F(r)\sqrt{1-\frac{r^2}{R^2}}}{k_0r^2}\!\right).
\label{rhoe}
\eeq
The functions $M(r)$ and $Q(r)$ defined in Eqs.~(\ref{massdef})
and (\ref{chargedef}), respectively, are then
\beq
M(r) =
\frac{r^3}{2R^2}+\frac{a}{2\left(2-a\right)^2}
\frac{k_0^2}{R^4}\frac{r^5}{
F^2(r) },
\label{mass-funct}
\eeq
\beq
Q(r) = \frac{\epsilon \sqrt{a\,}}{2-a}\frac{ k_0}{R^2}\,
\frac{r^3}{F(r)}\,.
\label{charge-sol1}
\eeq
The function $F(r)$ has the following definition
\beq
F(r) =  k_0\, \sqrt{1 - \frac{r^2}{R^2}}-k_1 \, ,\label{Fr}\\
\eeq
with the integration constants $k_0$ and $k_1$ given by
\beq
k_0 =\frac{R^2}{r_0^2}
\left(\frac{m}{r_0}-\frac{q^2}{r_0^2}\right) \left(1- \dfrac{r_0^2}
{R^2}\right)^{-1/a}\, ,
\label{constk}
\eeq
\beqa
k_1&=& \!\!k_0\,\sqrt{1-\dfrac{r_0^2}{R^2}}\!
\left[1 - \frac{a}{2-a}\!\dfrac{r_0^2}{R^2}
\left(\dfrac{m}{r_0}-\dfrac{q^2}{r_0^2}
\right)^{\!\!-1}\right], \quad
\label{constk_1}
\eeqa
where
\beq
m\equiv M(r_0)\,,
\label{m_1}
\eeq
\beq
q\equiv Q(r_0)\,.
\label{q_1}
\eeq
The constants $k_0$ and $k_1$ were found through the junction conditions. In 
this work we are interested in all possible values for the parameter $a$, 
$-\infty<a<\infty$. The limiting case $a \rightarrow\pm\infty$ yields the 
uncharged, $q=0$, Schwarzschild interior solution.

\subsubsection{Exterior solution}

For the external region, $r> r_0$, the solution of the Einstein-Maxwell field 
equations, Eqs.~\eqref{einsteq1}-\eqref{qspherical2}, 
is the Reissner-Nordstr\"om solution, i.e., 
\beq
{A(r)} =\frac{1}{1 -\dfrac{2m}{r}+ \dfrac{q^2}{r^2}}\,,
\label{ABext1}
\eeq
\beq
B(r) = \dfrac{1}{A(r)} = 1 -\dfrac{2m}{r}+ \dfrac{q^2}{r^2}\,,
\label{ABext2}
\eeq
\beq
\phi(r) = \dfrac{q}{r}\,, 
\label{phiext}
\eeq
\beq
\rho_{\rm m}(r)=0\,, 
\label{fluidext1}
\eeq
\beq
p(r)=0\,, 
\label{fluidext2}
\eeq
\beq
\rho_{\rm e}=0\,.
\label{rhoeext}
\eeq
The functions $M(r)$ and $Q(r)$ are simply
\beq
M(r) = m \,,  
\label{massext}
\eeq 
\beq
Q(r) = q \,,  
\label{chargeext}
\eeq
again, with $m$ and $q$ being the total mass and total charge of the exterior 
spacetime. By continuity, one finds that at $r=r_0$, the metric functions are 
$B(r_0)= 1/A(r_0)= 1 -{2m}/{r_0}+ {q^2}/{r_0^{2}}$. One must also have 
$\phi(r_0) = {q}/{r_0}$, $M(r_0)=m$, and $Q(r_0) =q$. Such a spherical 
electrovacuum spacetime has two radii associated with it, the gravitational 
and the Cauchy radii, given in terms of $m$ and $q$. The gravitational radius 
is given by 
\begin{equation}
 r_+ = m +\sqrt{m^2 - q^2}\,.
 \label{horizong}
\end{equation}
It is the horizon radius if there is a black hole.
The Cauchy radius is given by
\begin{equation}
 r_- = m - \sqrt{m^2 - q^2}\,. \label{horizonc}
\end{equation}

\subsubsection{Junction conditions}

The matching between the interior and the exterior solution is done at the 
boundary surface $r=r_0$. At this boundary the metric should be continuous. 
So, by making the junction of the interior metric function $g_{rr}=A(r)$ in 
Eq.~\eqref{A-sol1} to the $g_{rr}$ coefficient of the exterior metric function 
given by Eq.~\eqref{ABext2}, one finds \cite{guilfoyle} that
\beq
 m= \frac{r_0}{2}\left(\frac{r_0^2}{R^2} +
\frac{q^2}{r_0^2}\right).
\label{mass}
\eeq
An additional junction condition has to be imposed at $r=r_0$, namely, one 
must impose the continuity of the $g_{tt}=B(r)$ metric coefficient and the 
continuity of its first radial derivative to obtain
\beq
a=
\frac{r_0^2}{4q^2}\left(\frac{r_0^2}{R^2} -\frac{q^2}{r_0^2}\right)^2
\left(1-\frac{r_0^2}{R^2}\right)^{-1}. \label{a-function}
\eeq
where the fact that $Q(r_0)= q$ was also taken into account.
The constants $k_0$ and $k_1$ in Eqs.~(\ref{constk})-(\ref{constk_1}) 
have been also found through the junction conditions.

\subsubsection{Some constraints}
\label{sec-someconstraints}

If we restrict the solutions to be static spheres, the condition ${r_0} < {R}$ 
has to be obeyed. This condition has been imposed in~\cite{guilfoyle}.  On the 
other hand, when the matching surface $r_0$ is pushed to $R$ one finds regular 
black holes whose central core of charged matter fills up to the Cauchy 
horizon $r_0=r_- = R$~\cite{lemoszanchin2015}. Therefore, we impose the 
constraint
\begin{equation}
{r_0} \leq {R} \,. \label{r0R}
\end{equation}

In \cite{guilfoyle} it was displayed solutions satisfying some conditions, 
$m\geq |q|$ and $r_0 > r_+$, which implied $m> q^2/r_0$. Also some energy 
conditions were imposed. Here we release all these constraints, except 
Eq.~(\ref{r0R}), and look at the whole space of parameters. We give a physical 
interpretation to the resulting solutions. Clearly, there is a large region in 
the parameter space for which there are solutions, indeed there is a plethora 
of many interesting solutions.

\subsubsection{Further considerations }

Let us briefly mention some of the territory in the space of
solutions we are going to explore.

First, there is the special case $a=1$, not considered by Guilfoyle 
\cite{guilfoyle}. For the ansatz $B(r)= a \phi^2(r)$, Eq.~(\ref{wg-relation}), 
with $a=1$ one has the case with zero pressure, i.e., it is the 
Majumdar-Papapetrou ansatz for electrically charged fluids 
\cite{majumdar,papapetrou}. When this Majumdar-Papapetrou matter is joined 
into a spherically symmetric vacuum one gets the Bonnor stars with $q=m$ 
\cite{bonnorwickra2,lemoszanchin2005}.

Second, for $a\neq1$ and $a>0$, one finds from Eq.~(\ref{mass}) that there is 
the possibility   in which the equality $m\, r_0=q^2$ holds, which can yield, 
as a particular case, the relation $q=m$, an extremal solution. This case has 
also not been considered in  \cite{guilfoyle}. This limit, $m\, r_0=q^2$ and 
$q=m$ is the quasiblack hole limit \cite{lemoszanchin2010}. Since the 
inequality $a\neq 1$ holds in this case, it means the corresponding quasiblack 
hole is made of a fluid with nonzero pressure. As we will see, in fact, this 
point in the parameter space, namely, $q=r_0=m$, is degenerated.

Third, another region in the parameter space, also not explored in the 
original paper \cite{guilfoyle}, corresponds to extremely compact objects 
\cite{lemoszanchin2015}. These objects are found by allowing the central 
pressure to assume arbitrarily large values and the objects saturate the 
Buchdahl-Andr\'easson bound \cite{buchdahl,andreasson-charged}.

The above mentioned solutions are valid for all $a>0$, with the 
$a\rightarrow\infty$ needing a special treatment. The case $a=0$ is also 
special and requires a careful analysis in order to check if there are 
solutions in such a limit. If one allows $a<0$ then other interesting 
solutions may be found. In the following we study all the possibilities.

\section{General analysis, analysis of the parameters $\boldsymbol{m}$ and 
$\boldsymbol{a}$, and display of some special cases}
\label{sec-analysis}

\subsection{General analysis}
\label{genericsGA}

\subsubsection{The range of the parameters of the two-dimensional parameter 
space $q^2/R^2\times r_0/R$}
\label{genericsGAsub}

These solutions have five parameters, namely, $a$, $m$, $r_0$, $R$, and $q$.  
Taking into account the two relations \eqref{mass} and \eqref{a-function}, one 
can consider that the free parameters of the model are three, we choose them 
to be the electric charge $q$, the radius of the star $r_0$, and the energy 
density parameter $R$. In the respective parameter space, these parameters in 
principle have the following range: $0\leq r_0 < \infty$, $-\infty < q < 
\infty$, and $0\leq R <\infty$. The choice for the range of $r_0$ is obvious.  
The choice for the range of $q$ is based in the usual physical interpretation 
of the electric charge. However, we will do differently, and will allow $q$ to 
assume imaginary values, i.e., we will allow $q^2<0$. The interpretation of 
the parameter $R$, following its definition from Eq.~\eqref{rhoconst}, is that 
large $R$ means low energy densities while small $R$ represents large energy 
densities. The limiting value $R\rightarrow 0$ may be interpreted as a 
singular pointlike solution, while $R\rightarrow \infty$ represents empty 
space. To display in a clear way the spectrum of solutions we can do better, 
and without loss of generality, display the solutions in a two-dimensional 
parameter space instead of in the three-dimensional space spanned by $r_0$, 
$q$, and $R$.

For this let us give arguments to choose the two-dimensional dimensionless 
parameter space $q^2/R^2\times r_0/R$ as the most interesting choice and also 
give the range of $q^2/R^2$ and $r_0/R$.

We consider the dimensionless ratio $q^2/R^2$ as a first free parameter. Now, 
the constraint \eqref{r0R} together with Eqs.~\eqref{mass}-\eqref{a-function} 
imply $m$ and $a$ may be zero or negative only if $q^2$ is allowed to be 
negative. So, we may consider the dimensionless ratio $q^2/R^2$ with a range 
of $q^2/R^2$ not restricted, i.e., \begin{equation} 
-\infty<\frac{q^2}{R^2}<\infty\,. \label{q2R2} \end{equation} We are thus 
allowing for an imaginary electric charge. The range of $q^2$ is the same as 
the range of $q^2/R^2$. We use the usual convention for the modulus of the 
electric charge, i.e., for $q^2<0$ we denote $|q|=\sqrt{|q^2|}$ and for 
$q^2>0$ we denote $|q|=\sqrt{q^2}$.

We consider the dimensionless ratio $r_0/R$ as a second free parameter. The 
constraint \eqref{r0R} together with the size of the charged matter 
distributions, which is given by its radius $r_0$ and the total energy 
density, codified by the parameter $R$, are related to the compactness of the 
objects. For instance, large energy densities may be obtained by taking a 
fixed mass $m$ and diminishing $r_0$, or by  fixing $r_0$ and diminishing $R$. 
Moreover, the metric functions depend only on the ratio $r_0/R$, and, of 
course, on $r/R$. Hence, it is interesting to consider the ratio $r_0/R$ as 
the true free parameter of the model. This, together with the constraint 
$r_0\leq R$ imposed by the model, Eq.~(\ref{r0R}), implies that the range of 
the $r_0/R$ parameter is
\begin{equation}
0\leq \frac{r_0}{R}\leq 1 \,. \label{r0/R<1}
\end{equation}

The parameter space $q^2/R^2\times r_0/R$ is therefore two-dimensional. The 
range of the parameters given in Eqs.~\eqref{q2R2} and~\eqref{r0/R<1} defines 
then a strip in the two-dimensional parameter space. A preliminary analysis 
will be performed on the parameters $m$ and $a$ in terms of the free 
parameters  $q^2/R^2$ and $r_0/R$. Figures~\ref{curves-for-mass} to 
\ref{regions-detail-closer} will help in the understanding of the whole 
spectrum of solutions.

For the study that follows, it shall be convenient to divide the parameter 
space into two sectors: (I) 
$q^2/R^2 > 0$, and (II) $q^2/R^2<0$, the $q^2/R^2=0$ case 
being a vertical line. These sectors are properly indicated in 
Figs~\ref{curves-for-mass} and \ref{curves-for-a}.

\subsubsection{On the imaginary electric charge}
\label{imq}

From Eq.~\eqref{q2R2} the case of negative squared charge, $q^2/R^2  <0$ also 
enters the analysis. It deserves some comments.

Equation~\eqref{a-function} implies that for negative $q^2$ the parameter $a$ 
must be negative. This is clear after replacing $q^2$ by $-|q|^2$ and taking 
into account that $r_0^2/R^2$ is not larger than unity.

This has implications for the interior. From Eq.~\eqref{charge-sol1}
one finds that in this case, i.e., for $a<0$, $Q^2(r)$ is negative. So
$Q(r)$ is a pure imaginary number, as is $q$. The electric potential
$\phi$ is also a pure imaginary function to guarantee the metric
coefficient $B(r)$ is a real and positive function. On the other hand,
all the metric spacetime functions are real since they depend on $q^2$
which is a real negative parameter.

The negativity of $q^2$
has also implications for the exterior. For negative $q^2$ the metric 
function of the external Reissner-Nordstr\"om spacetime, $B(r) = {1}/{A(r)}=1 
-2m/r +q^2/r^2= 1 -2m/r-|q|^2/r^2$, Eq.~\eqref{ABext2}, has only one positive 
root, $r_+$, given by $ r_+ = m + \sqrt{m^2 + |q|^2} $, i.e., it is given as 
in Eq.~\eqref{horizong} with $q^2\to -|q|^2$. This is the gravitational radius 
of the solution, a horizon if $r_0<r_+$.  Using Eq.~\eqref{mass} we can show 
that for $q^2<0$ the gravitational radius $r_+$ is never larger than the  
radius of the matching boundary $r_0$, i.e., $ r_+ \leq r_0,$, for $q^2 <0$, 
with the equality holding in the limit $q^2 \rightarrow-\infty$. Hence, no 
black hole-type solutions are found for $q^2<0$. Further properties of
the solutions with $q^2< 0$ are discussed later.

For a negative $q^2$ it is not possible to interpret $q$ as electric charge. 
Notwithstanding, a negative $q^2$ term was suggested by Einstein and Rosen in 
considering singularity free massless particles \cite{einsteinrosen}. A 
similar situation happens for static spherical black hole solutions in a 
brane-world gravity scenario \cite{tidal}, where a tidal charge that may be 
negative plays the same role as the negative $q^2$ term in the 
Reissner-Nordstr\"om metric.

\subsection{Analysis of the mass parameter $\boldsymbol{m(q^2,r_0,R)}$}
\label{sec_mass}

\subsubsection{Generics}

The behavior of the mass $m$ is given in Eq.~\eqref{mass}. Clearly $m$ is a 
function of $q^2$, $r_0$ and $R$ (it is actually a function of $R^2$, but 
since $R>0$, the choice for $R$ or $R^2$ is indifferent, it is done here for 
reasons of better graphic displaying). Also from Eq.~\eqref{mass} one sees 
that the mass function $m(q^2,r_0,R)$ can be written formally as 
$m(q^2,r_0,R)=m(r_0,q^2/R^2,r_0/R)$. Figure~\ref{curves-for-mass} shows the 
behavior of the mass function $m(q^2,r_0,R)=m(r_0,q^2/R^2,r_0/R)$ given by 
Eq.~\eqref{mass}. Figures~\ref{regions-detail} and \ref{regions-detail-closer} 
will also be referred to in this analysis. 

To simplify the analysis, we consider separately the regions of the parameter 
space corresponding to five interesting classes of $m$: $m< 0$; $m=0$;  $0< m 
< |q|$; $m=|q|$; and $m > |q|$, where again, for $q^2<0$ we denote 
$|q|=\sqrt{|q^2|}$ and for $q^2>0$ we denote $|q|=\sqrt{q^2}$. Of special 
interest is the class $m=|q|$,  it is a curve that has two branches, namely, a 
closed curve in sector (I) and an open curve in sector (II), see 
Fig.~\ref{curves-for-mass}.

\subsubsection{The various classes of $m$}
\label{classesofm}

\vskip 0.2cm
\centerline{$m < 0$}

\begin{figure}[ht]
\begin{center}
\includegraphics[scale=1.1]{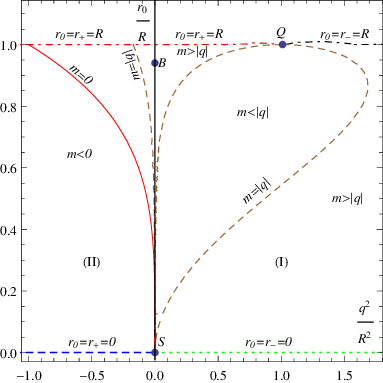}
\caption{Level curves for $m$ and corresponding areas in the $q^2/R^2\times 
r_0/R$ plane.
For $q^2/R^2>0$ the curve for $m=|q|$ coincides with the curve for $a=1$. Also 
indicated are the lines $r_0=r_-$ (dotted-dashed horizontal line $r_0/R =1$, 
with $q^2/R^2 \geq 1$), $r_0=r_+$ (dotted-dashed horizontal line $r_0/R =1$, 
with $ q^2/R^2 \leq 1$),  $r_+=r_0=0$, and $r_-=r_0 =0$, as well as the points 
$B$ representing the Buchdahl limit with $(q=0,m/r_0=4/9)$, $Q$ representing 
the quasiblack hole limit with $(r_+=r_-=r_0=m=|q|)$, and $S$ representing the 
Schwarzschild black hole limit with $(q=0,r_0=0)$.  }
\label{curves-for-mass}
\end{center}
\end{figure}

Since we are allowing mass contribution from the electromagnetic field to be
negative,  the total gravitational mass of the system can be negative. The
related region of the parameter space is below the curve $m=0$ as seen in
Fig.~\ref{curves-for-mass}. In Figs.~\ref{regions-detail} and
\ref{regions-detail-closer} the region corresponding to the $m<0$ solutions
is labeled by (h). 

A particular case that can be dealt with analytically is the case $m=-|q|$. 
This solution is obtained by putting $-q^2 = m^2$ into Eq.~\eqref{mass}. The 
resulting negative mass is given by $ m=r_0\left(-1- \sqrt{1 
+{r_0^2}/{R^2}}\right)$. The corresponding curve is not drawn in 
Fig.~\ref{curves-for-mass}. The curve would be in region (h) of 
Figs.~\ref{regions-detail} and \ref{regions-detail-closer} but it
is also not drawn.

All the solutions with $m< 0$ are singular and  seem to be of 
little physical interest.

\vskip 0.7cm
\centerline{$m=0$}

The condition
\begin{equation}
m(r_0,q^2/R^2,r_0/R)=0\,, \label{m=00}
\end{equation}
imposed to Eq.~\eqref{mass}
furnishes the relation
\begin{equation}
\frac{q^2}{R^2} = -\frac{r_0^4}{R^4}\,. \label{m=0}
\end{equation}
The Guilfoyle parameter is negative in this case, it is given by $a= 
-({r_0^2}/{R^2})\left(1-{r_0^2}/{R^2}\right)$. This yields the curve $m=0$ 
shown in sector (II) of Fig.~\ref{curves-for-mass}, see also 
Figs.~\ref{regions-detail} and \ref{regions-detail-closer} where the curve 
$m=0$ is denoted by $C_5$. The solutions on this line present singularities 
and have little importance as star models.

\vskip 0.7cm
\centerline{$0<m<|q|$}

The solutions with the mass $m$ in this range of values represent stars, see 
Fig.~\ref{curves-for-mass}. The star radius $r_0$ is larger than the 
gravitational radius, i.e., $r_0>r_+$ both when $q^2>0$ or $q^2<0$. For 
$q^2>0$ the stars are overcharged. Some of these solutions are regular, as 
those in regions (b) and (i) of Figs.~\ref{regions-detail} and 
\ref{regions-detail-closer}, while others are singular, as those in regions 
(c) and  (g) of the same figures. Notice that only parts of regions (g) and 
(i) contains object that satisfy $0<m<|q|$. The main  aspects of these 
positive mass overcharged starts are examined in some detail later.

\vskip 0.7cm
\centerline{$m=|q|$}

The condition 
\begin{equation}
m^2(r_0,q^2/R^2,r_0/R)=\pm q^2\,,\label{m2=q2}
\end{equation}
imposed to Eq.~\eqref{mass} yields two equations depending on the sign of 
$q^2$. We analyze the two cases separately.

\vskip 0.5cm
\noindent
$q^2/r_0^2<0$: In this case Eqs.~(\ref{mass}) and (\ref{m2=q2})
give
\begin{equation}
\frac{m^2}{r_0^2} + \frac{2m}{r_0} - \frac{r_0^2}{R^2} =0\,.
\label{m=q-}
\end{equation}
There are two solutions to this equation, namely, $
m=r_0\left(-1\pm \sqrt{1 +{r_0^2}/{R^2}}\right)$. One of them yields a
negative mass, which was already considered in the case $m<0$ above. 
Here we take the one that yields positive mass, 
\begin{equation}
m=|q|= r_0\left(-1 + \sqrt{1 +\frac{r_0^2}{R^2}}\right)\, ,
\label{massq2neg}
\end{equation}
or
\begin{equation}
m^2= -q^2=r_0^2\left(2+\frac{r_0^2}{R^2} - 2\sqrt{1 +
\frac{r_0^2}{R^2}}\right)\,. \label{q2R2negmpos}
\end{equation}
This equation 
generates the line indicated by the appropriate label in sector (II) of 
Fig.~\ref{curves-for-mass}.  The solutions on this line can be regular stars 
or singular objects,  and the pressure does not vanishes inside the fluid for 
these stars.  The regular stars are located in region (i), while the singular 
solutions are located in region (g) of Figs.~\ref{regions-detail} and 
\ref{regions-detail-closer}. These solutions have not been studied in the 
literature.  The region to the right-hand side of such a curve contains 
charged starlike solutions and other solutions with $m>|q|$, while the region 
to the left of that curve contains solutions with $m< |q|$.

\vskip 0.5cm
\noindent
$q^2/r_0^2>0$:
In this case Eqs.~(\ref{mass}) and (\ref{m2=q2}) give 
\begin{equation}
\frac{m^2}{r_0^2} -\frac{2m}{r_0} + \frac{r_0^2}{R^2} =0\,. \label{m=q+}
\end{equation}
The solution to this equation is 
\begin{equation} m=|q|=r_0\left(1\pm \sqrt{1 
- \frac{r_0^2}{R^2}}\right)\, , \label{pm2}
\end{equation}
or 
\begin{equation}
m^2= q^2=r_0^2\left(2-\frac{r_0^2}{R^2} \pm 2\sqrt{1 -
\frac{r_0^2}{R^2}}\right)\,. \label{q2R2pos}
\end{equation} 
The closed curve indicated by the label $m = |q|$ in the sector (I) of 
Fig.~\ref{curves-for-mass} is the complete solution, found joining the plus 
and minus solutions of Eq.~(\ref{pm2}). In Figs.~\ref{regions-detail} and 
\ref{regions-detail-closer} the $m = |q|$ solution is the sum of the curves 
denoted by $C_0$ and $C_2$. In the Guilfoyle's solutions \cite{guilfoyle} the 
condition $m = |q|$ is identical to the condition $a=1$.  This means that for 
this class of solutions the pressure is zero everywhere inside the fluid 
\cite{lemoszanchin2005} (see also \cite{lemoszanchin2009}), corresponding to 
charged dust stars and also to quasiblack holes \cite{lemoszanchin2010}.

\vskip 0.7cm
\centerline{$m>|q|$ }

This class has objects for a large range of parameters and in several
regions of the parameter space.

In the region $q^2/R^2 <0$, sector (II) in Fig.~\ref{curves-for-mass}, the 
solutions with $m>|q|$ are in the small portion of the sector between the axis 
$q^2/R^2=0$ and the line $m=|q|$. Objects in this region may be regular stars, 
as those in the part of region (i) in Figs.~\ref{regions-detail} and 
\ref{regions-detail-closer} above the curve $m=|q|$, or singular objects such 
those in the part of region (g) above the curve $m=|q|$.

In the region $0<q^2/R^2<1$, within sector (I) in Fig.~\ref{curves-for-mass}, 
the objects with $m>|q|$ are locate in the region above the curve $m=|q|$.  In 
Figs.~\ref{regions-detail} and \ref{regions-detail-closer} this region is 
labeled by (f) and (a). The Buchdahl-Andr\'easson bound divides the 
undercharged star solutions that are regular (in region (i)) from those that 
are singular (in region (f)). For the Buchdahl-Andr\'easson bound line see 
below and see curve $C_4$ in Figs.~\ref{regions-detail} and 
\ref{regions-detail-closer}).

In the region $1<q^2/R^2<\infty$, within sector (I) in 
Fig.~\ref{curves-for-mass}, the objects with $m>|q|$ are located are located 
in the region outside the curve $m=|q|$.  Here the constraint $r_0< r_-$ holds 
and the solutions are regular black holes, see~\cite{lemoszanchin2016} for 
more information. In Figs.~\ref{regions-detail} and 
\ref{regions-detail-closer} these solutions cover regions (d) and (e).

\subsection{Analysis of the parameter $\boldsymbol{a(q^2,r_0,R)}$}

\subsubsection{Generics}

\begin{figure}[ht]
\vskip -.0cm
\begin{center}
\includegraphics[scale=1.1]{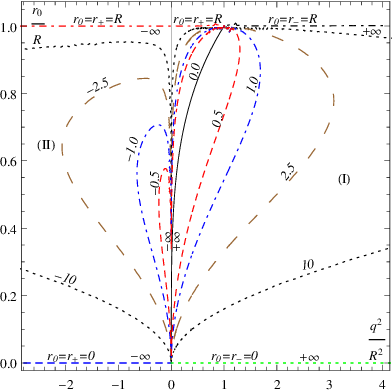}
\caption{Several contours for different values of the parameter $a$ as a
function of the normalized charge ($q^2/R^2$) and radius of the star
$r_0/R$.
The set of curves in the $q^2>0$ sector [sector (I)] are all for positive
values of $a$, including $a=0$.
The  set of curves in the $q^2<0$ sector [sector (II)] are all for negative 
values of $a$. Curves with the following $a$s are drawn: $a=\pm 0.0$ [solid 
line in sector (I) only], $a=\pm0.5$ (dashed closed line, in both sectors), 
$a=\pm 1.0$ (dotted-dashed closed line, in both sectors), $a=\pm 2.5$
(spaced-dashed closed line, in both sectors), $a=\pm 10$ (dotted external 
curve, in both sectors). 
The $a=\pm\infty$ curves are composed of the  three axes (top horizontal, 
vertical, bottom horizontal) in the $q^2\geq0$ sector, and the  three axes 
(top horizontal, vertical, bottom horizontal) in the $q^2\leq0$ sector, 
respectively, and can be thought as closing at $q^2/R^2\to\pm\infty$,
respectively. 
All contours are closed curves (the $a=0$ line can be considered a trivial
closed curve). See text for details.}
\label{curves-for-a}
\end{center}
\end{figure}

The behavior of the Guilfoyle parameter $a$ is given in 
Eq.~\eqref{a-function}. Clearly $a$ is a function of $q^2$, $r_0$ and $R$ (it 
is actually a function of $R^2$, but since $R>0$, the choice for $R$ or $R^2$ 
is indifferent, it is done here for reasons of better graphic displaying). 
Also from Eq.~\eqref{a-function} one sees that the Guilfoyle parameter 
$a(q^2,r_0,R)$ can be written formally as $a(q^2,r_0,R)=a(q^2/R^2,r_0/R)$. 
Figure~\ref{curves-for-a}, where some important curves are drawn, shows the 
behavior of the Guilfoyle parameter $a(q^2,r_0,R)=a(q^2/R^2,r_0/R)$ given by 
Eq.~\eqref{a-function}. We identify some special regions in the parameter 
space according to the values of $a$ and give the general properties of the 
solutions belonging to each one of such regions. Figures~\ref{regions-detail} 
and \ref{regions-detail-closer} will also be referred in this analysis. 
Figure~\ref{curves-for-mass}, which shows some curves for the mass $m$, is 
also useful in the study of the parameter $a$.

From Eq.~\eqref{a-function}, taking into account the parameter intervals 
given in Eqs.~\eqref{q2R2} and \eqref{r0/R<1}, it follows that $a$ assumes 
values in the interval
\begin{equation}
-\infty< a<\infty\,.
\label{intervalofa}
\end{equation}
As seen from Eq.~\eqref{a-function}, the limit of infinite $a$, $a\to\infty$, 
corresponds to a vanishing electric charge. Also, from 
Eq.~\eqref{wg-relation}, in this limit the electric potential vanishes in such 
a way that the geometric function $B(r)$ remains finite. In this sense the 
Schwarzschild interior solution is a particular case of the Guilfoyle's 
solutions. The limit $a\to-\infty$ has the same properties as the limit 
$a\to\infty$. Moreover, $a$ also acquires arbitrarily large values when 
$r_0/R=0$ for $q^2\neq 0$, and $r_0/R =1$ for all values of $q^2/R^2$ except 
$q^2/r_0^2 =1$. In the other limit, the limit of vanishing $a$, $a=0$, it 
requires that the electric charge assumes arbitrarily large values, in such a 
way that the limit of $a\phi^2 $ is finite. This case is considered here for 
completeness. The parameter $a$ also vanishes for $q^2/R^2 = r_0^4/R^4$ with 
$r_0^2/R^2\neq 1$.  Hence, $a=0$ is also of interest in the present study. The 
Guilfoyle parameter is therefore in the range given in Eq.\eqref{intervalofa}.

Now it is convenient to consider separately the two sectors: (I) $q^2/R^2 > 
0$ and $0\leq r_0/R \leq 1$, where $a$ is positive; (II) $q^2/R^2 < 0$ and 
$0\leq r_0/R \leq 1$, where $a$ is negative, as indicated in 
Fig.~\ref{curves-for-a}, see also Fig.~\ref{curves-for-mass}. We draw some 
curves for constant values of $a(q^2/R^2,r_0/R)$ for the ranges given in 
Eqs.~\eqref{q2R2} and \eqref{r0/R<1}, i.e., $-\infty<q^2/R^2 <\infty$ and 
$0\leq r_0/R \leq 1$, see Fig.~\ref{curves-for-a}. Considering that the 
$a=-\infty$ line closes at $q^2/R^2\to-\infty$, that $a=0$ line is a closed 
curve, actually trivially closed, and that the $a=-\infty$ line closes at 
$q^2/R^2\to\infty$, one finds that all contours $a={\rm constant}$ are closed 
curves.

To simplify the analysis, we study separately the regions of the parameter 
space in seven interesting classes of $a$: $a=-\infty$; $-\infty<a< 0$; $a=0$; 
$0 < a<1$;  $a =1 $;  $1<a<\infty$;  $a=\infty$. Below we investigate each one 
of these classes.

\subsubsection{The various classes of $a$}

\vskip 0.2cm
\centerline{\it $ a = -\infty$: Singular solutions}
\centerline{\it and uncharged Schwarzschild stars}

The limit $a=-\infty$ arises only for $q^2\leq0$. In Fig.~\ref{curves-for-a}, 
the $a=-\infty$ limit is composed of the top horizontal line $r_0/R=1$ with 
$r_0=r_+=R$ in the sector $q^2<0$, plus the  vertical line 
$q^2/r_0^2=0$, plus the bottom horizontal line $r_0/R=0$  with $r_0=r_-=0$ in 
the sector $q^2<0$. One can then think that the top and bottom horizontal lines 
join to form a closed curve at $q^2/R^2\to-\infty$. Indeed, the $a=-\infty$ 
limit appears in several instances that can be deduced using 
Eq.~\eqref{a-function} as we now show.

One instance is when $r_0/R\rightarrow 1$. Here, one finds that the 
$a=-\infty$ limit admits charged singular solutions with $r_0/R=1$  and where 
$r_0=r_+=R$, see Fig.~\ref{curves-for-a}.

Another instance is for $q^2/r_0^2\rightarrow 0$ with nonzero $r_0/R$, i.e., 
the zero charged solutions with finite nonzero radius $r_0$ yield $a = 
-\infty$. These solutions compose the interior Schwarzschild solution and 
correspond to the vertical line $q^2/R^2=0$ in Fig.~\ref{curves-for-a}.

Yet another instance is when $r_0/R\rightarrow 0$. Here, one finds
that the $a=-\infty$ limit admits charged singular solutions where
$r_0=r_+=0$, see Fig.~\ref{curves-for-a}.

Finally, one can then think that the top and bottom horizontal lines
join at $q^2/R^2\to-\infty$ to form a closed curve.

\vskip 0.7cm
\centerline{\it ${-\infty<a<0}$: Regular stars and singular solutions}

The negative $a$ class, ${-\infty<a<0}$, includes all the sector (II) in 
Fig.~\ref{curves-for-a}, which covers the parameter space portion given by 
$q^2/R^2<0$ (i.e., imaginary electric charge) and $0<r_0/R <1$. In 
Fig.~\ref{curves-for-a} four representative curves of this class, namely, 
$a=-10$, $a=-2.5$, $a=-1.0$, and $a=-0.5$, are shown as a dotted closed
line, a spaced-dashed closed line, and a dotted-dashed closed line, and a
dashed closed line, respectively.

In this class,
the matching surface $r_0$ is outside the gravitational radius $r_+= m + 
\sqrt{m^2 + |q|^2}$ (see Eq.~\eqref{horizong} and the discussion in 
Sec.~\ref{imq}). So there are no black holes in this class. The solutions can 
be regular or singular, depending on the relative values of the parameters. A 
more complete discussion of sector (II) is given below, see also 
Figs.~\ref{regions-detail} and \ref{regions-detail-closer}.

\vskip 0.7cm
\centerline{\it ${a=0}$: The Schwarzschild vacuum solution,}
\centerline{\it the quasiblack hole, and generic singular solutions}

The curve 
\begin{equation} a(r_0,q^2/R^2, r_0/R)=0, \label{a=00}
\end{equation} 
is shown in Fig.~\ref{curves-for-a}, it is the solid line in region (I). In 
Figs.~\ref{regions-detail} and \ref{regions-detail-closer} the curve is 
indicated by $C_1$. Considering the Guilfoyle relation, 
Eq.~\eqref{wg-relation}, one sees that Eq.~\eqref{a=00}, will give solutions 
with $B(r)=0$. Note also that from Eq.~\eqref{a-function} the condition $a=0$ 
only occurs for $q^2/R^2>0$, i.e., $q^2>0$. Bearing in mind this restriction 
we can say the following for $a=0$.

From Eq.~\eqref{a-function}, one sees that Eq.~\eqref{a=00}, is satisfied in 
the plane $q^2/R^2\times r_0/R$, by the curve
\begin{equation} \frac{r_0}{R} = 
\sqrt{\frac{|q|}{R}}, \label{a=0}
\end{equation}
for $q^2/R^2\neq 0$ and $r_0/R \neq 0$,  $q^2/R^2\neq  1$ and $r_0/R \neq 1$, 
and $q^2/R^2>0$ as stated above. The corresponding mass is $m= 
r_0^3/R^2=q^2/r_0$. The solutions in this curve represent singular solutions. 
They are without physical interest except for the two limiting end points of 
the curve.

The two limiting points of the curve, namely, $q^2/R^2= 0$ and $r_0/R =0$, and 
$q^2/R^2= 1$ and $r_0/R =1$, must be analyzed with care. In the case 
$q^2/R^2=0$ and $r_0/R= 0$, one finds that to get  $a=0$ the point must be 
approached in a specific way, namely, from Eq.~\eqref{a-function} one should 
put first $r_0^4=R^2q^2$, and only then one takes $r_0^2/R^2\to0$ and so 
$q^2/r_0^2\to0$, then also $a=0$ (if one had put $q^2/r_0^2=0$ without 
$r_0^4=R^2q^2$ holding, then from Eq.~\eqref{a-function} one  gets 
$a\to\infty$). This point is the point $S$ displayed in 
Figs.~\ref{regions-detail} and \ref{regions-detail-closer} and through this 
limit $S$ represents the  Schwarzschild vacuum solution. The other possible 
case with Eq.~\eqref{a=00} holding is the other limiting point $q^2/R^2=1$ and 
$r_0/R= 1$. It is the quasiblack hole point $Q$ also  displayed in 
Figs.~\ref{regions-detail} and \ref{regions-detail-closer}. These points will 
be discussed later on.

\vskip 0.7cm
\centerline{\it $0<a<1$: Overcharged tension stars}
\centerline{\it and singular solutions}

The class for which the condition $0 < a < 1$ holds occupies a portion in 
sector (I) of Fig.~\ref{curves-for-a}. It is also the same portion for which 
$m < |q|$, with $q^2>0$, see Fig.~\ref{curves-for-mass}.  In 
Fig.~\ref{curves-for-a} one representative curve of this class, namely, 
$a=0.5$, is shown as a dashed closed line. In Figs.~\ref{regions-detail} and 
\ref{regions-detail-closer} such a portion covers the regions indicated by (b) 
and (c), and also the curve  $C_1$. The matching surface is outside the 
gravitational radius, $r_0>r_+$, so the solutions in this portion are not 
black holes. For relatively large values of $r_0^2/q^2$, namely, for $r_0^4> 
q^2R^2$, within this portion, the solutions are regular everywhere, whereas 
for $r_0^4< q^2R^2$, the solutions are singular in the sense that there are 
surfaces of infinite pressure or infinite energy density for some radius $r$ 
inside $r_0$. In Figs.~\ref{regions-detail} and \ref{regions-detail-closer}, 
region (b) contains overcharged tension (negative pressure) stars, region (c) 
contains the singular solutions just mentioned.

\vskip 0.7cm
\centerline{\it $a=1$:  Charged dust stars and other solutions}

The equation 
\begin{equation} a(q^2/R^2,r_0/R)=1, \label{a=11}
\end{equation} 
defines a closed curve indicated by a dotted-dashed line in  sector (I) of 
Fig.~\ref{curves-for-a}. It is also the closed curve labeled as $m=|q|$
for $q^2/R^2$ 
in 
Fig.~\ref{curves-for-mass}. In Figs.~\ref{regions-detail} and 
\ref{regions-detail-closer} this closed curve is given by the union of  the 
curves $C_0$ and $C_2$.

All the solutions on this curve have zero pressure, $p(r)=0$, and the matter 
is extremely charged $\rho_{\rm m}=\rho_{\rm e}$, i.e., it is made of 
Majumdar-Papapetrou matter. For all these solutions,
the  spacetime region exterior to the boundary
$r>r_0$ is the extreme Reissner-Nordstr\"om spacetime, for which $m=|q|$.

The left part of the $a=1$ curve in Fig.~\ref{curves-for-a}, labeled by $C_0$ 
in Figs.~\ref{regions-detail} and \ref{regions-detail-closer}, contains
solutions that represent zero pressure regular charged stars.  These are the
Bonnor stars studied in the literature. The right part of the $a=1$ curve
in Fig.~\ref{curves-for-a}, labeled by $C_2$ in Figs.~\ref{regions-detail}
and \ref{regions-detail-closer}, contains solutions that represent zero
pressure singular solutions, the energy density $\rho_{\rm m}$
reaching arbitrarily large values for certain values of the radius $r$
within the star.

The two parts of the $a=1$ curve  meet at two points, the upper point $Q$
and the lower point $S$. The point $Q$ is the quasiblack hole limit 
(see Figs.~\ref{regions-detail} and \ref{regions-detail-closer}). It is a 
degenerated point in the sense that, besides solutions with zero pressure, it 
contains solutions with nonzero pressure. $Q$ is the quasiblack hole 
point~\cite{lemoszanchin2010}. The point $S$ is also degenerated, it 
represents different solutions. The limit along these two parts of the curve 
gives the Minkowski empty spacetime. More details on these points are given 
later on.

\vskip 0.7cm
\centerline{\it $ 1<a<\infty$: Charged stars, regular black holes,}
\centerline{\it singular  black holes, and other singular solutions}

The class for which the condition  $ 1<a<\infty$ holds occupies an outer 
portion in sector (I) of Fig.~\ref{curves-for-a}. In Fig.~\ref{curves-for-a} 
two representative curves of this class, namely, $a=2.5$ and $a=10$, are shown 
as a spaced-dashed closed line and a dotted closed line, respectively.

There are several distinct subclasses of solutions. These subclasses appear
distinctively in Figs.~\ref{regions-detail} and \ref{regions-detail-closer},
and the corresponding regions are indicated by (a) which has undercharged
stars studied in the original work by Guilfoyle \cite{guilfoyle}, (d) which
shows regular black holes with negative energy density
\cite{lemoszanchin2016}, (e) which presents regular black holes with a
central core of phantom matter studied in~\cite{lemoszanchin2016},  and (f)
which has singular solutions. A description of the properties of each class
of solutions will be done later.

\vskip 0.7cm
\centerline{\it $a=\infty$: Regular black holes with a de Sitter core,}
\centerline{\it singular solutions, and uncharged Schwarzschild stars}

The limit $a=\infty$ arises only for $q^2\geq0$. In Fig.~\ref{curves-for-a}, 
the $a=\infty$ limit is composed of the top horizontal line $r_0/R=1$ with 
$r_0=r_+=R$ and $r_0=r_-=R$ in the sector $q^2>0$, plus the  vertical line 
$q^2/r_0^2=0$, plus the bottom horizontal line $r_0/R=0$  with $r_0=r_-=0$ in 
the sector $q^2>0$. One can then think that the top and bottom horizontal line 
join to form a closed curve at $q^2/R^2\to\infty$. Indeed, the $a=\infty$ 
limit appears in several instances that can be deduced using 
Eq.~\eqref{a-function} as we now show.

One instance is when $q^2/r_0^2 \neq r_0^2/R^2$ and $r_0/R\rightarrow 1$. One  
finds that the $a=\infty$ limit admits charged solutions with $r_0/R=1$ for 
which the matching radius $r_0$ coincides with the inner horizon $r_-$ of the 
Reissner-Nordstr\"om black hole \cite{lemoszanchin2016} (see also 
\cite{lemoszanchin2011}). These are regular black holes containing a perfect 
fluid of de Sitter type and a coat of electric charge at the surface boundary 
$r_0$, where $r_0=r_-=R$, see Fig.~\ref{curves-for-a}. One also finds that the 
$a=\infty$ limit admits charged singular solutions with $r_0/R=1$  and where 
$r_0=r_+=R$, see Fig.~\ref{curves-for-a}. Note that the excluded point in the 
analysis above, $q^2/R^2=1$, $r_0/R=1$, is the quasiblack hole point, 
indicated by $Q$ in Figs.~\ref{regions-detail} and 
\ref{regions-detail-closer}. It is characterized by $r_0=r_-=r_+=R$. In this 
case $a(q,r_0)$ is not defined there, it has any value including $a=\infty$, 
depending on how each one of the parameters approaches unity. A more detailed 
analysis is given later.

Another instance is for $q^2/r_0^2\rightarrow 0$ with nonzero $r_0/R$, i.e., 
the zero charged solutions with finite nonzero radius $r_0$ yield $a = 
\infty$. These solutions compose the interior Schwarzschild solution and 
correspond to the vertical line $q^2/R^2=0$ in Fig.~\ref{curves-for-a}. 

Yet another instance is when $r_0/R\rightarrow 0$. Here, one finds that the 
$a=\infty$ limit admits charged singular solutions where $r_0=r_-=0$, see 
Fig.~\ref{curves-for-a}.

Finally, one can then think that the top and bottom horizontal lines join at 
$q^2/R^2\to\infty$ to form a closed curve.

\subsection{Special nontrivial cases that can be displayed analytically}

\subsubsection{The Schwarzschild
interior solution and the Schwarzschild star}
\label{schwint}

In the limit of zero electrical charge $q^2=0$ one obtains expressions for the 
metric potentials and for the fluid quantities directly from the Guilfoyle's
solution. In this limit $q^2=0$ one also has that the $a$ parameter obeys 
$a=\pm\infty$, see Eq.~\eqref{a-function}. Taking then the limit $q^2 
\rightarrow 0$ one gets for the functions given
in  Eqs.~\eqref{A-sol1}-\eqref{rhoe} the following expressions, 
\beq
A(r) =\left({1 - \dfrac{r^2}{R^2}}\right)^{-1}\, , \label{ASchwint}
\eeq
\beq
B(r) =
\frac{1}{4}\left(3\sqrt{1-\frac{r_0^2}{R^2}}
-\sqrt{1-\frac{r^2}{R^2}}\right)^2, \label{BSchwint}
\eeq
\beq
\phi(r)=0, \label{phiSchw} 
\eeq
\beq
8\pi\rho_{\rm m}(r) = \frac{3}{R^2}, \label{rhomSchw} 
\eeq
\beq
8\pi p(r)= -1 + \frac{2 \sqrt{1 - \dfrac{r^2}{R^2}}}{3\sqrt{1 -
\dfrac{r_0^2}{R^2}}-\sqrt{1 -\dfrac{r^2}{R^2}}}, \label{prSchw}
\eeq
\beq
\rho_{\rm e} (r) = 0\,. \label{rhoeSchw}
\eeq
The functions $M(r)$ and $Q(r)$ of Eqs.~\eqref{mass-funct} 
and~\eqref{charge-sol1} are now 
\beq
M(r) = \frac{1}{2}\frac{r^3}{R^2}, \label{massSchw} 
\eeq
\beq
Q(r) = 0\, \label{qSchw} 
\eeq
respectively.
The auxiliary function $F(r)$ given in Eq.~\eqref{Fr} is still
\beq
F(r)= k_0\, \sqrt{1 - \frac{r^2}{R^2}}-k_1 \,,
\label{frschw} 
\eeq
and $k_0$ and $k_1$  given in Eqs.~\eqref{constk} and  \eqref{constk_1} are 
now
\beq
k_0=\frac{m}{r_0}\frac{R^2}{r_0^2}, \label{k00} 
\eeq
\beq
k_1=\left(1+\frac{m}{r_0}\frac{R^2}{r_0^2}\right)
\sqrt{1-\dfrac{r_0^2}{R^2}}, \label{k11} 
\eeq
respectively. The
auxiliary function $F(r)$ and $k_0$ and $k_1$ 
are already embedded into Eqs.~\eqref{ASchwint}-\eqref{rhoeSchw}.
Also the mass $m$ and the electric charge $q$ defined in Eqs.~\eqref{m_1}
and~\eqref{q_1} are now
\beq
m = M(r_0)= \frac{1}{2}\frac{r_0^3}{R^2}, \label{massSchw11} 
\eeq
\beq
q = Q(r_0) =0. \label{qSchw11} 
\eeq
Of course this set of equations are the Schwarzschild interior solution. The 
exterior solution is the Schwarzschild spacetime, with the metric coefficients 
of Eqs.~\eqref{ABext1} and \eqref{ABext2}
being now given by 
\beq
{A(r)} =\frac{1}{1 -\dfrac{2m}{r}}\,,
\label{ABextschw11}
\eeq
\beq
B(r) = \dfrac{1}{A(r)} = 1 -\dfrac{2m}{r}\,.
\label{ABextsvh22}
\eeq
These metric potentials should then be inserted
into the metric~\eqref{metricsph}. The electric 
potential and the electric field components are equal to zero. Interior and 
exterior Schwarzschild solutions
compose what one may call the Schwarzschild star.

Note that from Eq.~\eqref{massSchw} the mass $m$ of these Schwarzschild stars 
goes to zero with $r_0^3$. Therefore, the resulting solution in the limit of 
$r_0=0$ is the Minkowski spacetime.

In Figs.~\ref{curves-for-mass}-\ref{regions-detail-closer} the Schwarzschild 
stars are contained in the line segment $q^2/R^2=0$ and $0< r_0/R <1$. When 
one takes the limit of the line $q^2/R^2 = 0$ down to $ r_0/R =0$ one arrives 
at the point  $S$ shown in Fig.~\ref{curves-for-mass} and 
Figs.~\ref{regions-detail}-\ref{regions-detail-closer}, which in this limit 
represents Minkowski spacetime. In Fig.~\ref{curves-for-a} the line 
$q^2/R^2=0$ and $0< r_0/R <1$ is part of the curves $a\rightarrow\pm\infty$.

\subsubsection{The Buchdahl-Andr\'easson bound}
\label{babound}

An important relation between the parameters of the solutions, i.e., between 
$m$, $q$, and $r_0$, appears when the fluid pressure at the center of the star 
goes to infinity.
This limit gives the Buchdahl-Andr\'easson bound~\cite{lemoszanchin2015},
\begin{equation}
\frac{m}{r_0}= {\left(\frac{1}{3}+\sqrt{\frac{1}{9}+\frac{q^{2}}{3r_0^{2}
} } \right)^{2}}\,. \label{bab}
\end{equation}

From Eq.~\eqref{bab} one can extract three interesting cases. For 
$q^2/r_0^2=0$, i.e., the zero electrically charged case, one gets the Buchdahl 
bound, $m/r_0=4/9$~\cite{lemoszanchin2015}. For $q^2/r_0^2=1$, i.e., the 
extremal case, one gets the quasiblack hole limit, $m/r_0=1$ 
\cite{lemoszanchin2010}. For $q^2/r^2= -1/3$, i.e., the maximal compact star, 
one gets the maximum value of the Buchdahl-Andr\'easson bound,  $m/r_0 = 9$, 
indeed a very compact star that has negative $q^2$, not considered 
in~\cite{lemoszanchin2015}, but clearly the bound also holds for negative 
$q^2$ as long as one has $q^2/r_0^2 \geq - 1/3$.

The Buchdahl-Andr\'easson bound appears as  the curve $C_4$ in 
Figs.~\ref{regions-detail} and \ref{regions-detail-closer}.

\subsubsection{The case $r_0=r_+$}
\label{r0=r+}

When the radius of the object $r_0$ reaches $r_+$ one has two
situations. From Eq.~(\ref{horizong}) one has $r_0= 
m(q,r_0,R)+\sqrt{m^2(q,r_0,R)-q^2} $ and from Eq.~(\ref{mass}) this yields
\begin{equation}
r_0= \frac{q^2}{2r_0} +\frac{r_0^3}{2R^2} +
\sqrt{\left(\frac{q^2}{2r_0} +\frac{r_0^3}{2R^2}\right)^2 -q^2}\,.
\label{radiuseq+}
\end{equation}
Equation~(\ref{radiuseq+}) has two solutions.

One solution is 
\begin{equation}
r_0=r_+ = R\,,
\label{r0=r+=R}
\end{equation}
which is satisfied  as long as $-\infty< q^2/R^2 < 1$.
It represents infinite pressure solutions with matching at
the Reissner-Nordstr\"om horizon, i.e., singular solutions.

The other solution is 
\begin{equation}
r_0=r_+ = 0\,,
\label{r0=r+=0}
\end{equation}
satisfied for all $q^2\leq 0$. It represents negative mass naked 
singularities, see Appendix~\ref{1} for details.

In Figs.~\ref{curves-for-mass}-\ref{regions-detail-closer} these two sets of
singular solutions are represented and explicitly indicated by the
appropriate labels. In Fig.~\ref{curves-for-a} the solutions mentioned in
this section are part of the curves $a=\pm\infty$.

\subsubsection{The case $r_0=r_-$}
\label{r0=r-}

When the radius of the object $r_0$ reaches $r_-$ one has very
interesting situations. From Eq.~(\ref{horizonc}) one has $r_0= 
m(q,r_0,R)-\sqrt{m^2(q,r_0,R)-q^2} $ and from Eq.~(\ref{mass}) this yields 
\begin{equation} r_0=\frac{q^2}{2r_0} +\frac{r_0^3}{2R^2} - 
\sqrt{\left(\frac{q^2}{2r_0} +\frac{r_0^3}{2R^2}\right)^2 -q^2}\,. 
\label{radiuseq-} 
\end{equation} 
Equation~(\ref{radiuseq-}) has two solutions.

One solution is 
\begin{equation} r_0 =r_-= R\,, \label{r0=r-=R} 
\end{equation} 
which is satisfied only for $q^2/R^2> 1$. It gives regular black holes with
a de Sitter core with a matching at the Cauchy horizon. These solutions were
studied in~\cite{lemoszanchin2016} (see also \cite{lemoszanchin2011}).

The other solution is
\begin{equation} r_0 =r_-= 0\,, \label{r0=r-=0} 
\end{equation} 
satisfied for all $q^2/R^2 > 0$. It is a Kasner spacetime time, see
Appendix~\ref{2} for details.

In Figs.~\ref{curves-for-mass}-\ref{regions-detail-closer} these two sets of
solutions are represented and explicitly indicated by the appropriate
labels. In Fig.~\ref{curves-for-a} the solutions mentioned above are part of
the curves $a=\pm\infty$.

\subsubsection{The case $r_0=r_+=r_-$}
\label{r0=r+=r-}

When the radius of the object $r_0$ reaches a double horizon $r_+=r_-$ one has 
from~Eqs.~(\ref{horizong})-(\ref{mass}) the result 
\begin{equation} 
r_0=r_+=r_-=m=q=R\,. \label{radiuseq+-} 
\end{equation} 
This is the quasiblack hole. It is an extremal object with very interesting 
properties, for details see \cite{lemoszanchin2010}.

In Figs.~\ref{curves-for-mass}-\ref{regions-detail-closer} the quasiblack
hole is represented. In Fig.~\ref{curves-for-mass} and
Figs.~\ref{regions-detail}-\ref{regions-detail-closer} it is explicitly
indicated by the appropriate labels. In Fig.~\ref{curves-for-a} it is a
point in all positive $a$ curves, in particular in the curve $a=\infty$.

\subsection{Generics to the next two sections}

A point in the parameter space $q^2/R^2\times r_0/R$ represents one solution. 
Thus all solutions are represented in this parameter space. These solutions 
can be physical or unphysical, although this division can be subjective. In 
the sector $q^2\geq0$, the physical ones are normal undercharged stars, dust 
extremely charged stars, overcharged tension stars, and regular black holes 
with a phantom matter core, and the unphysical ones are regular black holes 
with negative energy densities and singular solutions.  In the sector $q^2<0$, 
there are regular and singular solutions.

Figures~\ref{regions-detail} and \ref{regions-detail-closer} display the 
regions, areas in the figure, where the solutions are regular, represented by 
white regions, and solutions where the pressure and, or, the energy-density 
inside the fluid distribution are singular, represented by gray regions. The 
boundaries of these regions are lines or points which also represent 
solutions. In the next two sections we comment carefully what the regions and 
boundaries represent physically. We start with regions in the $q^2/R^2\geq0$ 
sector [regions (a), (b), (c), (d), (e), and (f)], some of them have solutions 
with interesting physical meaning, and then continue into regions in the 
$q^2/R^2<0$ sector [regions (g), (h), and (i)]. We also study the
lines 
and points, boundaries of the analyzed areas, namely, the curves $C_0$,
$C_1$,
$C_2$,
$C_3$,
$C_4$,
$C_5$,
the horizontal and vertical axes and lines, and the points $S$, $B$, and $Q$.

The analysis has been performed using Mathematica and exploring in
detail the  whole spectrum of this plethora of solutions. It was a
meticulous work.

\section{Regular and singular regions (areas) in the 
parameter space $\boldsymbol{q^2/R^2\times r_0/R}$}

\label{sec-regions}

\subsection{Region (a): Undercharged stars}

\begin{figure}[ht]
\vskip -.1cm
\begin{center}
\includegraphics[scale=1.1]{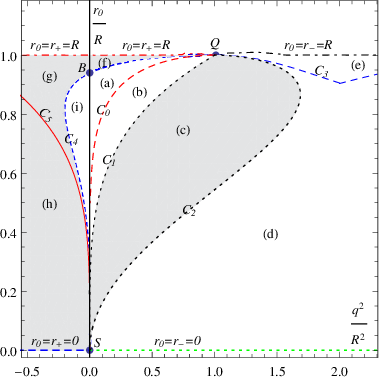}
\caption{The white regions contain regular  solutions and
the gray regions contain singular solutions. The
boundaries between two regular regions contain regular solutions. The
boundaries between a regular region and a singular one, or between two
singular regions, contain singular solutions. }
\label{regions-detail}
\end{center}
\end{figure}

\begin{figure}[ht]
\vskip -.1cm
\begin{center}
\includegraphics[scale=1.1]{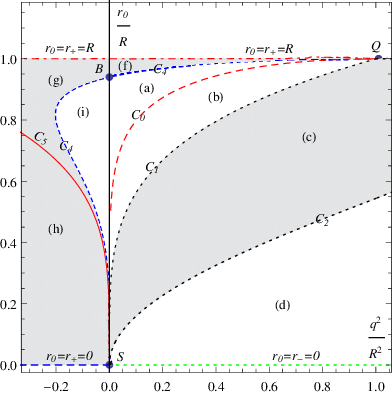}
\caption{The same as Fig.~\ref{regions-detail} zoomed in. Regions with
regular (white) and singular (gray) solutions and their boundaries are
indicated.}
\label{regions-detail-closer}
\end{center}
\end{figure}

Region (a) is the region of undercharged nonsingular stars and is delimited by 
three curves, namely, the $q^2/R^2=0$ coordinate axis from the origin up to 
the point $B$, the sector of curve $C_4$ joining $B$ to the point $Q$ 
(separating gray and white regions), and the curve of zero pressure indicated 
by $C_0$, see Figs.~\ref{regions-detail} and \ref{regions-detail-closer}.  
$C_0$ is a branch of the curve $m=|q|$, or $a=1$, and $C_4$ is the 
Buchdahl-Andr\'easson curve, see also Figs.~\ref{curves-for-mass} 
and~\ref{curves-for-a}.

Some more detail is now given. 
Undercharged stars are characterized by $m>|q|$. In this nonsingular region of 
undercharged stars, region (a), the constraint $r_0/r_+ > 1$ is satisfied, so 
that black holes are not present. Moreover, the Guilfoyle parameter $a$ 
satisfies the constraint $a> 1$ in the whole region, except at the boundaries. 
The stars with radius $r_0$ and charge $q$ belonging to this region of the 
parameter space satisfies the energy conditions, as studied by Guilfoyle 
\cite{guilfoyle}. Further analysis of the objects, including the point $Q$ in 
Figs.~\ref{regions-detail} and \ref{regions-detail-closer}, is given in 
\cite{lemoszanchin2009}.

The boundaries of region (a) have solutions whose  physical properties
are discussed later.

\subsection{Region (b): Overcharged tension stars}

Region (b) is the region of overcharged nonsingular tension stars and is 
delimited by $C_0$ and $C_1$ in Figs.~\ref{regions-detail} and 
\ref{regions-detail-closer}. $C_0$ is a branch of the curve $m=|q|$, or $a=1$, 
and $C_1$ is the curve given by $a=0$, see also Figs.~\ref{curves-for-mass} 
and~\ref{curves-for-a}.

Some more detail is now given. 
Overcharged stars are characterized by $m<|q|$. The parameter space for the
whole set, singular and nonsingular overcharged solutions, with $m<|q|$ can
be found using Eq.~\eqref{mass}. The condition  $m\leq|q|$ gives
$|q|^2/r_0^2 -2|q|/r_0 + r_0^2/R^2 \leq0$, see also Eq.~(\ref{m=q+}). The
equality furnishes the curve  $m=|q|$ which for $q^2/R^2 > 0$, i.e.,
$q^2>0$, is the closed curve formed by the $C_0$ and $C_2$ curves in
Figs.~\ref{regions-detail} and \ref{regions-detail-closer}. Thus, the
overcharged region of the parameter space is the whole region inside such a
curve. The curve $C_1$ separates the nonsingular from the singular
overcharged solutions. Indeed, in the $C_1$ line the Guilfoyle parameter
vanishes, $a(q^2,r_0)=0$ [see Eq.~\eqref{a=0}] and the fluid quantities are
singular. So, nonsingular overcharged solutions are delimited by $C_0$ and
$C_1$.

Note that the curve $m=|q|$ in the sector $q^2/R^2 > 0$ is the closed curve
explicitly given in Fig.~\ref{curves-for-mass}. Since $m=|q|$ coincides
with the $a=1$ curve, one also finds that the overcharged region corresponds
to small values of the Guilfoyle parameter $a$, namely, $0 < a < 1$, and
that in these solutions the pressure within the fluid is negative $p(r) < 0$
for all solutions there, i.e., it is a tension.

The boundaries of region (b) have solutions whose  physical properties are 
discussed later.

\subsection{Region (c): Overcharged solutions, wildly behaved
solutions}
\label{region_e}

Region (c) is the region of overcharged singular solutions with wild behavior 
of the energy density and pressure and is delimited by the curves $C_1$ and 
$C_2$, see Figs.~\ref{regions-detail} and \ref{regions-detail-closer}. $C_1$ 
is the curve given by $a=0$, and $C_2$ is a branch of the curve $m=|q|$, or 
$a=1$, see also Figs.~\ref{curves-for-mass} and~\ref{curves-for-a}.

Some more detail is now given. These are overcharged solutions still with 
$m<|q|$. The energy density of the solutions in this region behaves wildly, it 
is positive and finite at the center and changes to very high negative values 
close to the surface. The pressure of these solutions behaves as wildly as the 
energy density, it is negative and finite at the center, and oscillates to 
very high positive values close to the surface of the solution and finally 
goes to zero at the surface. Because of this wild behavior, these solutions 
are highly weird and we do not study them further.

The boundaries of region (c) have solutions whose  physical properties are 
discussed later.

\subsection{Region (d): Regular black holes with negative energy
densities}

Region (d) is the region of regular black holes with negative energy delimited 
by the curves $C_2$ and $C_3$, and the horizontal line $r_0=r_-=0$ in 
Fig.~\ref{regions-detail}, part of it being seen also in 
Fig.~\ref{regions-detail-closer}. $C_2$ is a branch of the curve $m=|q|$, or 
$a=1$,  $C_3$ is a curve with no special features in terms of the parameters 
$m$ or $a$, and the horizontal line $r_0=r_-=0$ is a portion of the curve 
$a=\infty$, see also Figs.~\ref{curves-for-mass} and~\ref{curves-for-a}.

Some more detail is now given. The solutions in region (d) are characterized 
by several facts. First, the matching surface defined by $r_0$ is timelike, 
and its radius is smaller than the Cauchy horizon radius $r_-$, $r_0/r_-< 1$. 
So the solutions represent black holes. Second, the solutions are regular, 
i.e., there are no singularities. Third, the energy density is negative for a 
range of radii in the solutions and the pressure can also be negative going to 
zero at the surface $r_0$. Fourth, due to the negativity of the energy 
density, these regular black holes have less interest than those of region (e) 
below.

The boundaries of region (d) have solutions whose  physical properties are 
discussed later.

\subsection{Region (e): Regular black holes with a phantom 
matter core}
\label{region_c}

Region (e) is the region of regular black holes with a phantom matter core and 
is delimited by the curve $C_3$ and the line $r_0=r_-=R$ in 
Fig.~\ref{regions-detail}, it is outside the range of 
Fig.~\ref{regions-detail-closer}. $C_3$ is a curve with no special features in 
terms of the parameters $m$ or $a$, the line $r_0=r_-=R$ is a branch of the 
$a=\infty$ curve, see also Figs.~\ref{curves-for-mass} and~\ref{curves-for-a}.

Some more detail is now given.  The solutions in region (e), like those in
region (d) above, are characterized by several facts. First, the matching
surface defined by $r_0$ is timelike, and its radius is smaller than the
Cauchy horizon radius $r_-$, $r_0/r_-< 1$. So the solutions represent black
holes. Second, the solutions are regular, i.e., there are no singularities. 
Third, the energy density is positive everywhere inside matter, the pressure
is negative and it goes to zero at the surface $r_0$.  Fourth, this negative
pressure at the center is larger than the central density, thus for a finite
region inside the matter one finds $\rho_{\rm m}+ p<0$, meaning there is
phantom matter and consequently the energy conditions are violated.  In
brief, in region (e) one gets regular black holes with an electrically
charged phantom matter core. These regular black holes were investigated
in~\cite{lemoszanchin2016}.

The boundaries of region (e) have solutions whose  physical properties are 
discussed later.

\subsection{Region (f): Singular solutions}

Region (f) is a region of singular solutions,  delimited by three curves, 
namely, the $q^2/R^2=0$ coordinate axis from the point $B$ up to $r_0/R=1$, by 
the curve $C_4$ in the $q^2/R^2>0$ sector, and by the line $r_0=r_+=R$, see 
Figs.~\ref{regions-detail} and \ref{regions-detail-closer}. The curve $C_4$ is 
the Buchdahl-Andr\'easson bound curve, and the vertical line $q^2/R^2=0$ and 
the horizontal line $r_0=r_+=R$ are part of the $a=\infty$ curve, see also 
Figs.~\ref{curves-for-mass} and~\ref{curves-for-a}.

All solutions in region (f) are singular, since the pressure diverges
at some point inside the matter distribution. 

The boundaries of region (f) have solutions whose  physical properties
are discussed later.

\subsection{Region (g): Singular solutions with $\boldsymbol{m 
>0}$ and $\boldsymbol{q^2<0}$}
\label{region_j}

Region (g) is the region of singular solutions with $m>0$ and $q^2/R^2<0$, 
i.e., $q^2<0$, and is delimited by the $q^2/R^2=0$ coordinate axis from the 
point $B$ up to $r_0/R=1$, the horizontal line $r_0=r_+=R$, the curve  $C_5$, 
and 
the curves $C_4$  in the sector $q^2/R^2<0$, see Fig.~\ref{regions-detail} and 
in more detail Fig.~\ref{regions-detail-closer}. The curve $C_4$ is the 
Buchdahl-Andr\'easson bound curve, the vertical line $q^2/R^2=0$ and the 
horizontal line $r_0=r_+=R$ are a portion of the curve $a=-\infty$, and the 
curve  $C_5$ is the curve $m=0$, see also Figs.~\ref{curves-for-mass} 
and~\ref{curves-for-a}.

For all solutions in this region of the parameter space the energy density 
becomes infinitely large and the pressure becomes arbitrarily large and 
negative for some radius inside the matter distribution. Thus, this region
of the parameter space bears solutions of limited physical interest.

The boundaries of region (g) have solutions whose  physical properties are 
discussed later.

\subsection{Region (h): Singular solutions with 
$\boldsymbol{m < 0}$
and $\boldsymbol{q^2<0}$}
\label{region_k}

Region (h) is the region of singular solutions with negative mass, $m<0$, 
which implies here $q^2/R^2<0$, i.e., $q^2<0$, and is delimited by the curve 
$C_5$ and the horizontal line $r_0=r_+=0$, see Fig.~\ref{regions-detail} and 
in more detail Fig.~\ref{regions-detail-closer}. The curve $C_5$ is the curve 
$m=0$ and the horizontal line $r_0=r_+=0$ is a portion of the curve 
$a=-\infty$, see also Figs.~\ref{curves-for-mass} and~\ref{curves-for-a}.

This region presents solutions which are similar to those belonging to
region (g), the only difference is that here the mass is negative, $m<0$.
The solutions are singular, the matter interior is plagued with energy
densities arbitrarily large, and the pressure also assumes arbitrarily large
negative values. These singular solutions have also limited interest.

The boundaries of region (h) have solutions whose physical properties are 
discussed later.

\subsection{Region (i): Regular stars with $\boldsymbol{m>0}$
and $\boldsymbol{q^2<0}$}
\label{region_g}

Region (i) is the region of regular stars with positive mass, $m>0$, and 
$q^2/R^2<0$, i.e., $q^2<0$, and is delimited by the $q^2/R^2=0$ coordinate 
axis from the origin up to the point $B$, and a portion of the curve $C_4$ in 
the sector $q^2/R^2<0$, see Figs.~\ref{regions-detail} 
and~\ref{regions-detail-closer}. The vertical line $q^2/R^2=0$ is part of the 
curve $a=-\infty$ and the curve $C_4$ is the Buchdahl-Andr\'easson bound 
curve, see also Figs.~\ref{curves-for-mass} and~\ref{curves-for-a}.

Some more detail is now given. In this region since $q^2<0$ the electric 
charge $q$ is imaginary, and the Guilfoyle parameter $a$ is negative, $a<0$. 
One finds that the radius $r_0$ of the matching surface is greater than the 
gravitational radius, $r_0>r_+$ where $r_+=m +\sqrt{m^2+|q|^2}$. The solutions 
are nonsingular. So the solutions represent charged stars with imaginary 
electric charge.

The boundaries of region (i) have solutions whose  physical properties are 
discussed later.

\section{
Regular and singular boundaries (lines and points) in the 
parameter space $\boldsymbol{r_0/R\times q^2/R^2}$ }
\label{sec-boundaries}

\subsection{Curve $\boldsymbol{C_0}$: Bonnor stars} 

The curve $C_0$ is the line of charged dust stars. It begins at point $S$ and 
extends to point $Q$, see Figs.~\ref{regions-detail} and 
\ref{regions-detail-closer}. $C_0$  corresponds to $m=|q|$ in the sector 
$q^2>0$ and is part of the closed $a=1$ curve, see also 
Figs.~\ref{curves-for-mass} and~\ref{curves-for-a}.

On $C_0$ the fluid pressure vanishes throughout the interior of the matter
distribution, so it is dust, and the spacetime region exterior to the boundary 
$r>r_0$ is the extreme $m=|q|$ Reissner-Nordstr\"om spacetime. The stars in 
this curve are called Bonnor stars, i.e., charged dust stars  
\cite{bonnorwickra2,lemoszanchin2008}.

\subsection{Curve $\boldsymbol{C_1}$: Boundary of regular overcharged
tension star solutions}

The curve  $C_1$ has overcharged solutions and it separates region (b) of 
regular overcharged stars with tension from region (c) of overcharged singular 
solutions with tension, see Figs.~\ref{regions-detail} and 
\ref{regions-detail-closer}. $C_1$ has objects with $m<|q|$ in the sector 
$q^2>0$ and it corresponds to $a=0$, i.e., it is the curve given by the 
equation $q^2/R^2 = r_0^4/R^4$, see Eq.~(\ref{a=0}), see also 
Figs.~\ref{curves-for-mass} and~\ref{curves-for-a}.

On $C_1$ the solutions are singular.

\subsection{Curve $\boldsymbol{C_2}$: Singular charged dust solutions}

The curve $C_2$ is the line of singular charged dust solutions, and as $C_0$  
it begins at point $S$ and extends to point $Q$, see 
Figs.~\ref{regions-detail} and \ref{regions-detail-closer}. $C_2$ also  
corresponds to $m=|q|$ in the sector $q^2>0$ and is the other part of the 
closed $a=1$ curve, see also Figs.~\ref{curves-for-mass} 
and~\ref{curves-for-a}.

On $C_2$ the solutions contain matter that is extremely charged,  the pressure 
is zero, however, the energy density is negative assuming very large values 
close to the center of the solution, a similar behavior to the solutions of 
region (c). We regard these solutions as singular solutions.

\subsection{Curve $\boldsymbol{C_3}$: Boundary between
regular black holes with phantom matter
and regular black holes with negative energy matter}

The curve $C_3$ is the line that divides regular black holes with negative 
energy matter shown in region (d) from regular black holes with phantom matter 
shown in region (e), see Figs.~\ref{regions-detail} 
and~\ref{regions-detail-closer}. The curve $C_3$ has no relation with the $m$ 
and $a$ curves displayed in Figs.~\ref{curves-for-mass} 
and~\ref{curves-for-a}.

For the regular black holes on $C_3$, the matching surface radius $r_0$ is 
smaller than the Cauchy horizon of the Reissner-Nordstr\"om metric $r_-$, 
$r_0< r_-$.  Even though being regular, the matter functions inside $r_0$ 
present some peculiar properties, for instance, the energy density vanishes 
for some radii $r$.

\subsection{Curve $\boldsymbol{C_4}$: The Buchdahl-Andr\'easson bound}
\label{babound1}

The curve $C_4$ is the Buchdahl-Andr\'easson bound line, i.e., it is the line 
obtained taking the limit of the central pressure going to infinity, see 
Figs.~\ref{regions-detail} and~\ref{regions-detail-closer}. The curve $C_4$ 
has no relation with the $m$ and $a$ curves displayed in 
Figs.~\ref{curves-for-mass} and~\ref{curves-for-a}.

The Buchdahl-Andr\'easson bound is discussed in Sec.~\ref{babound}, 
particularly Eq.~(\ref{bab}). The Buchdahl-Andr\'easson bound penetrates the 
region of negative electrical charge, $q^2<0$, a region not considered 
in~\cite{lemoszanchin2015}. The bound holds as long as one has $q^2/r_0^2 \geq 
- 1/3$. The maximum value of the Buchdahl-Andr\'easson bound is $m/r_0 = 9 $ 
for $q^2/r_0^2 = -1/3$, a very compact star with negative $q^2$. The curve 
$C_4$ contains the three very special and important points, the quasiblack 
hole point $Q$ ($q^2/R^2=1$, $r_0/R=1$), the Buchdahl point $B$ ($q^2/R^2=0$, 
$r_0/R=2\sqrt{2}/3$), and the Schwarzschild point $S$ ($q^2/R^2=0$, 
$r_0/R=0$).

\subsection{Line $\boldsymbol{r_0=r_+=R}$ and $\boldsymbol{0< q^2 <
R^2}$: Singular solutions with the
matter boundary at the Reissner-Nordstr\"om
gravitational radius}
\label{ror+R1}

The line  $r_0/R=r_+/R=1$ for $0< q^2/R^2 < 1$ is a line that represents 
singular solutions with the matter boundary at the Reissner-Nordstr\"om 
gravitational radius $r_0/r_+=1$, it starts at $q^2/R^2=0$ and goes up to 
point $Q$, see Figs.~\ref{regions-detail} and~\ref{regions-detail-closer}. 
This line has no relation with the $m$ curves and is a part of the $a=\infty$ 
curve, see also Figs.~\ref{curves-for-mass} and~\ref{curves-for-a}.

Each point on this line represents a solution where the interior region, i.e., 
from $r=0$ up to $r=r_0=r_+$, is filled with a charged fluid, and it is 
matched to the exterior Reissner-Nordstr\"om region by a lightlike surface. 
However, this matching can be realized only for infinite pressure, and thus 
the solutions are not well defined, they are singular. The line is discussed 
in  Sec.~\ref{r0=r+}.

\subsection{Line $\boldsymbol{r_0=r_-=R}$ and
$\boldsymbol{q^2>R^2}$: Regular black holes with a de
Sitter core and the matter boundary at
the Cauchy horizon}

The line $r_0/R=r_-/R=1$ for $q^2/R^2 >1$ is a line that represents regular 
black holes with a de Sitter core and the matter boundary at the Cauchy 
horizon 
$r_0/r_-=1$, it starts at point $Q$ and goes up to infinity to the right, see 
Figs.~\ref{regions-detail} and~\ref{regions-detail-closer}. This line has no 
relation with the $m$ curves and is a part of the $a=\infty$ curve, see also 
Figs.~\ref{curves-for-mass} and~\ref{curves-for-a}.

Each point on this line represents a regular charged black hole whose interior 
region is de Sitter, the exterior is Reissner-Nordstr\"om, and the matching 
surface is a layer of uniform charge density, the total charge being located 
on the boundary surface, i.e., at the Cauchy horizon. These solutions were 
studied in detail in~\cite{lemoszanchin2016} (see also 
\cite{lemoszanchin2011}).  The line is discussed in Sec.~\ref{r0=r-}.

\subsection{Line $\boldsymbol{r_0=r_-=0}$ and $\boldsymbol{q^2> 0}$:
Kasner spacetimes} \label{liner0=0+}

The line $r_0/R=r_-/R=0$ for $q^2/R^2 >0$ is a line for which each point 
represents a Kasner spacetime, see Figs.~\ref{regions-detail} 
and~\ref{regions-detail-closer}. This line has no relation with the $m$ curves 
and is a part of the $a=\infty$ curve, see also Figs.~\ref{curves-for-mass} 
and~\ref{curves-for-a}.

The Kasner spacetime has planar symmetry and appears in
several contexts, see Appendix~\ref{2}.

\subsection{Line $\boldsymbol{0< r_0 <
R}$ and $\boldsymbol{q^2=0}$: The Schwarzschild star}
\label{q20line}

The line $0< r_0/R <1$ for $q^2/R^2 =0$ is a line that represents the whole 
class of neutral, i.e., electrically uncharged, solutions within the 
Guilfoyle's solutions~\cite{guilfoyle}. It starts at the point $S$ and goes up 
vertically up to the point $r_0/R =1$, see Figs.~\ref{regions-detail} 
and~\ref{regions-detail-closer}. This line has no relation with the $m$ curves 
and is a part of both the $a=\infty$ and the $a=-\infty$ curves, see also 
Figs.~\ref{curves-for-mass} and~\ref{curves-for-a}.

Each point on this line represents a Schwarzschild star, i.e., the 
Schwarzschild interior solution matched to the  Schwarzschild exterior 
solution, see Sec.~\ref{schwint}. Note that from Eq.~\eqref{massSchw} the mass 
of these stars goes to zero as $r_0^3$. Therefore, the resulting solution 
found by taking the limit to the point $S$ down the line $q^2/R^2=0$ is the 
Minkowski spacetime, see Figs.~\ref{regions-detail} and 
\ref{regions-detail-closer}.

\subsection{Curve $\boldsymbol{C_5}$: Singular zero mass solutions }

The curve  $C_5$ represents zero mass solutions which can only exist in these 
solutions for $q^2 <0$, see Figs.~\ref{regions-detail} 
and~\ref{regions-detail-closer}. It also appears in Fig.~\ref{curves-for-mass} 
as the curve $m=0$ in the $q^2<0$ sector, in sector (I). This line has no 
relation with the $a$ curves, see Fig.~\ref{curves-for-a}.

The Guilfoyle parameter $a$ is negative for these solutions. The fluid
quantities in this curve are not well defined and the solutions are of
no physical interest. See also Sec.~\ref{classesofm},
particularly Eq.~(\ref{m=0}).

\subsection{Line $\boldsymbol{r_0=r_+=R}$
and $\boldsymbol{q^2< 0}$: Singular solutions with the
matter boundary at the Reissner-Nordstr\"om
gravitational radius in the negative electric charge square sector }

The line  $r_0/R=r_+/R=1$ for $q^2/R^2 < 0$ is a line that represents singular 
solutions with the matter boundary at the Reissner-Nordstr\"om gravitational 
radius, it starts at $q^2/R^2=-\infty$ and goes up to point $q^2/R^2=0$, see 
Figs.~\ref{regions-detail} and~\ref{regions-detail-closer}. This line has no 
relation with the $m$ curves, see Fig.~\ref{curves-for-mass}, and is a part of 
the $a=-\infty$ curve, see Fig.~\ref{curves-for-a}.

Each point on the line $r_0=r_+=R$, $q^2<0$ represents a solution where the 
interior region, i.e., from $r=0$ up to $r=r_0=r_+$, is filled with a charged 
fluid with negative square charge, and it is matched to the exterior 
Reissner-Nordstr\"om region by a lightlike surface. However, this matching can 
be realized only for infinite pressure, and thus the solutions are not well 
defined they are singular. These solutions have similar behavior to those with 
$r_0=r_+=R$, $0<q^2<R^2$ discussed in Sec.~\ref{ror+R1}. The line is also 
discussed in Sec.~\ref{r0=r+}.

\subsection{Line $\boldsymbol{r_0=r_+=0}$ and $\boldsymbol{q^2< 0}$:
Naked singularities in the negative electric charge square sector}
\label{liner0=0-}

The line  $r_0/R=r_+/R=0$ for $q^2/R^2 < 0$ is a line that represents naked 
singularities, see Figs.~\ref{regions-detail} and~\ref{regions-detail-closer}. 
This line has no relation with the $m$ curves, see Fig.~\ref{curves-for-mass}, 
and is a part of the $a=-\infty$ curve, see Fig.~\ref{curves-for-a}.

In this case the spacetime for small $r$ has planar symmetry and is singular 
at $r=0$, see Sec.~\ref{r0=r+}. It represents naked singularities, see also
Appendix~\ref{1}.

\subsection{Point S, $\boldsymbol{q^2 =0,\, r_0 = 0}$: The Minkowski
spacetime, the Kasner solution, the Schwarzschild solution,
the Schwarzschild negative mass singularity,
and the plane-symmetric static
spacetime naked singularity}

\subsubsection{Point S from above, $q^2 =0,\, r_0
\rightarrow 0$:  Minkowski spacetime}
\label{Sq20r0+}

When one approaches point $S$ from above, i.e., performing $q^2/R^2 =0$ and 
$r_0/R\to 0$, see Figs.~\ref{regions-detail} and \ref{regions-detail-closer}, 
one finds a Minkowski spacetime.

In this limit the mass $m$ indeed vanishes and the result is the Minkowski 
spacetime as follows from Eq.~\eqref{massSchw} and we have commented in 
Sec.~\ref{q20line}. In addition, in this limit the parameter $a$ becomes 
arbitrarily large.

\subsubsection{Point S from the right, $r_0=r_-=0, \, q^2
\rightarrow 0^+$:  Kasner spacetime}
\label{q2=0+}

When one approaches point $S$ from the right, i.e., performing  
$r_0/R=r_-/R=0$ and $q^2/R^2 \rightarrow 0^+$, see Figs.~\ref{regions-detail}  
and \ref{regions-detail-closer}, one finds a Kasner spacetime.

The limiting process is to take $r_0/R$ to zero, independently of $q^2/R^2$, 
and then taking $q^2/R^2$ to zero. The first part of the process leads to the 
Kasner metric, and the second part of the process does not change the 
resulting geometry, because the charge and mass are transformed way by 
changing the coordinates. The result is then a Kasner spacetime, see also 
Sec.~\ref{liner0=0+} and Appendix~\ref{2}. In this limit $a=\infty$.

\subsubsection{Point S from the skew right,
${r_0>0\,, q^2 > 0}$: Minkowski spacetime, Schwarzschild black
hole, and Kasner spacetime}
\label{pointS}

\centerline{\it\small Generics}

When one approaches point $S$ from a skew path on the right, see 
Figs.~\ref{regions-detail} and \ref{regions-detail-closer}, one finds, 
depending on the approaching path, a Minkowski spacetime, or a Schwarzschild 
black hole, or a Kasner spacetime. All the three cases have zero electric 
charge and zero electric potential, of course, since point $S$ has $q^2=0$. 
Some physical quantities on $S$ have values that depend on how the point is 
approached. A particular important quantity is the mass. That is a reason to 
get different spacetimes on approaching $S$ by different paths.

To see this, let $\gamma$ be a path in the two-dimensional parameter
space spanned by the parameters $q^2/R^2$ and $r_0/R$ approaching $S$.
The curve $\gamma$ may be parametrized by $q^2/R^2$ so that along
$\gamma$ we may write $r_0/R = f(q^2/R^2)$.  To simplify the notation
from now onward define $x\equiv q^2/R^2$ and $y\equiv r_0/R$, so that
the two-dimensional space is $x\times y$ and along $\gamma$ we may
write $y=f(x)$ with $\gamma$ being parametrized by $x$. Assume $f$ to
be a piecewise continuous function of its argument.  Since $f(x)$ has
also to satisfy the condition $\displaystyle{\lim_{x\rightarrow
0}f(x)}=0$ in order to get to the point $S$ at the end of the limiting
process, for small $x$ it must be of the form $f(x) \sim x^\beta$ with
positive $\beta$.  Let $f^\prime(x)$ denote the derivative of $f(x)$
with respect to its argument. Then, three situations can happen. We
discuss them now.

\vskip 0.7cm
\centerline{\it\small Slow approach to S: Minkowski spacetime}

If $\displaystyle{\lim_{x\rightarrow 0}xf^\prime(x)/f(x) <1}$, then
$y=f(x)\sim \alpha x^{1-\delta}$ (i.e., $r_0/R \sim
\left(q^2/R^2\right)^{1-\delta})$, with positive $\alpha$ and $\delta$.
Moreover, the mass goes as $m \sim x^\delta/2\alpha$ and vanishes in the
limit $x\rightarrow 0$, so that the result in this case is the Minkowski
spacetime. Finally, the Guilfoyle parameter $a$ is zero if $\delta > 1/2$,
is finite if $\delta = 1/2$, and diverges if $\delta <1/2$.

\vskip 0.7cm
\centerline{\it\small Linear approach to S: Schwarzschild black hole }

If $\displaystyle{\lim_{x\rightarrow 0}xf^\prime(x)/f(x) =1}$, then the
function $y=f(x)$ tends to zero as $y=f(x)\sim\alpha x$ (i.e., $r_0/R \sim
\alpha q^2/R^2$), with positive $\alpha$. In this case,
$\displaystyle{\lim_{x \rightarrow 0}m}= \alpha^{-1}/2$. Hence,
$\displaystyle{\lim_{x\rightarrow 0}r_+}= 2m=\alpha^{-1}$  and
$\displaystyle{\lim_{x\rightarrow 0}r_-=0}$. The metric functions are $B(r)
= 1/A(r) = 1 - 2m/r$ for all $r > 0$, the electric
charge $q$ and the electric
charge function $Q(r)$ vanish, and so does the electric potential
$\phi(r)$.  In such a
case the result is a Schwarzschild black hole spacetime. The Guilfoyle
parameter $a$ diverges.

\vskip 0.7cm
\centerline{\it\small Fast approach to S: Kasner spacetime}

If $\displaystyle{\lim_{x\rightarrow 0}xf^\prime(x)/f(x) >1}$, then 
$y=f(x)\sim \alpha x^{1+\delta}$ (i.e., $r_0/R \sim 
\left(q^2/R^2\right)^{1+\delta})$, with positive $\alpha$ and $\delta$. The 
mass goes as $m \sim 1/(2\alpha\, x^\delta)$ and diverges in the limit 
$x\rightarrow 0$. This is exactly what happens when approaching the point $S$ 
along the line $r_0 =0,\, q^2>0$, and the result in this case is a Kasner 
spacetime as shown in Sec.~\ref{q2=0+} and in Appendix~\ref{2}. The Guilfoyle 
parameter $a$ diverges.

\vskip 0.7cm
\centerline{\it\small Synopsis}

As we have seen, for $q^2/R^2>0$, the mass function $m(q^2,r_0,R)$ may have 
different values at point $S$, the value it assumes there depends on how the 
point $S$ is approached. The limiting mass value may be zero, implying a 
Minkowski spacetime, it may be finite and nonzero, implying a Schwarzschild 
black hole, and it may diverge, implying a Kasner spacetime.

\subsubsection{Point S from the left, ${r_0=r_+=0, \, q^2
\rightarrow 0^-}$:  Kasner spacetime}
\label{r0q2=0-}

When one approaches point $S$ from the left, i.e., performing $r_0/R=r_-/R=0$ 
and $q^2/R^2 \rightarrow 0^-$, see Figs.~\ref{regions-detail} and 
\ref{regions-detail-closer}, one finds a Kasner spacetime. So the resulting 
spacetime is the same as found in Sec.~\ref{q2=0+}, by approaching point $S$ 
from the right performing $r_0/R=r_-/R=0$ and $q^2/R^2 \rightarrow 0^+$.

The limiting process is to take $r_0/R$ to zero, independently of $q^2/R^2$,
with the constraint $q^2/R^2<0$, and then taking $q^2/R^2$ to zero. The
first part of the process leads to a naked singularity metric with negative
mass, and the second part of the process makes the charge and mass be
transformed way by changing the coordinates. The result is then a naked planar
singularity, more precisely, it is a static Kasner spacetime, see 
Sec.~\ref{liner0=0-}, Appendix~\ref{1} and Appendix~\ref{2}.

\subsubsection{Point S from the skew left,
${r_0>0, q^2 < 0}$:  Minkowski spacetime, Schwarzschild negative mass
singularity, and a plane-symmetric static
spacetime naked singularity}
\label{pointS3}

\centerline{\it\small Generics}

When one approaches point $S$ from a skew path on the left, see 
Figs.~\ref{regions-detail} and \ref{regions-detail-closer}, one finds, 
depending on the approaching path, a Minkowski spacetime, or a Schwarzschild 
black hole, or a Kasner spacetime. All the three cases have zero electric 
charge and zero electric potential, of course, since point $S$ has $q^2=0$.

The analysis of Sec.~\ref{pointS} is also appropriate here, since the behavior 
of the mass function when reaching the point $S$ from negative values of 
$q^2<0$ is similar as for the $q^2>0$ case. We define here $x\equiv |q|^2/R^2$ 
and $y\equiv r_0/R$. Let $\gamma$ be a a path in the two-dimensional parameter 
space spanned by $x$ and $y$. Along $\gamma$ we may write $y=f(x)$ with 
$\gamma$ being parametrized by $x$. Assume $f$ to be a piecewise continuous 
function of its argument.  Since $f(x)$ has also to satisfy the condition 
$\displaystyle{\lim_{x\rightarrow 0}f(x)}=0$ in order to get to the point $S$ 
at the end of the limiting process, for small $x$ it must be of the form $f(x) 
\sim x^\beta$ with positive $\beta$.  Let $f^\prime(x)$ denote the derivative 
of $f(x)$ with respect to its argument. Then again, three situations can 
happen. We discuss them now.

\vskip 0.7cm
\centerline{\it\small Slow approach to S: Minkowski spacetime}

If $\displaystyle{\lim_{x\rightarrow 0}xf^\prime(x)/f(x) <1}$, then $f(x)\sim 
x^{1-\delta}$ (i.e., $r_0/R \sim \left(|q|^2/R^2\right)^{1-\delta})$, with 
positive $\alpha$ and $\delta$, the mass goes as $m \sim -x^\delta/2\alpha$, 
so is negative, and vanishes in the limit $x\rightarrow 0$. The result in this 
case is the Minkowski spacetime.

\vskip 0.7cm
\centerline{\it\small Linear approach to S: Schwarzschild negative mass
singularity}

If $\displaystyle{\lim_{x\rightarrow 0}xf^\prime(x)/f(x) =1}$, then the 
function $f(x)$ tends to zero as $y=f(x) \sim  \alpha x$ (i.e., $r_0/R \sim 
\alpha q^2/R^2$), with positive $\alpha$, and we have $\displaystyle{\lim_{x 
\rightarrow 0}m}= -\alpha^{-1}/2$, so the mass is negative. There are no 
horizons. The metric functions result in $B(r) = 1/A(r) = 1 + 2|m|/r$ for all 
$r > 0$, the electric charge $q$ and the electric charge function $Q(r)$ 
vanish, and so does the electric potential $\phi(r)$. In such a case there are 
no horizons and the result is a Schwarzschild negative mass singularity.

\vskip 0.7cm
\centerline{\it\small Fast approach to S: A plane-symmetric static}
\centerline{\it\small 
spacetime naked singularity}

If $\displaystyle{\lim_{x\rightarrow 0}xf^\prime(x)/f(x) >1}$, then the
function $f(x)$ tends to zero as $y=f(x)\sim \alpha x^{1+\delta}$ (i.e.,
$r_0/R \sim \alpha \left(|q|^2/R^2\right)^{1+\delta}$), with positive
$\alpha$ and $\delta$, and the mass goes as $m \sim -1/\left(2\alpha\,
x^\delta\right)$ which is negative and diverges in the limit $x\rightarrow
0$. This is exactly what happens along the line $r_0 =0,\, q^2>0$, but now
the mass is negative, so the metric assumes the form
$
 ds^2 = -{2|m|}\, dt^2/r + {r}\,dr^2/\left(2|m|\right) + r^2 d\Omega^2  $.
One can then transform this metric away into a  more common
plane symmetric static naked singularity metric form, more precisely, it
is a static Kasner spacetime, see 
 Sec.\ref{liner0=0+} and  Appendix~\ref{1}.

\vskip 0.7cm
\centerline{\it\small Synopsis}

As we have seen, for $q^2/R^2<0$, the mass function $m(q^2,r_0,R)$ assumes
at the point $S$ different values depending on how the point $S$ is
approached. In terms of the resulting spacetime, the limit is three-folded.
The limiting mass value may be zero, implying the Minkowski spacetime, may
be finite and nonzero, implying a negative mass naked Schwarzschild
singularity, and may diverge, resulting in a planar static naked singularity
Kasner type spacetime.

\subsection{Point B: The Buchdahl bound}

Point $B$ is the Buchdahl bound \cite{buchdahl}, here represented by a 
Schwarzschild star (i.e., a Schwarzschild interior solution matched to the 
Schwarzschild vacuum solution) for which $r_0/m =9/4$, see 
Figs.~\ref{regions-detail} and~\ref{regions-detail-closer}.

Point $B$ is the zero electric charge, i.e., $q^2=0$, limit of the 
Buchdahl-Andr\'easson bound curve $C_4$, see Sec.~\ref{babound}, particularly 
Eq.~(\ref{bab}), and Sec.~(\ref{babound1}) (for a detailed study of the 
Buchdahl-Andr\'easson bound see~\cite{lemoszanchin2015}). The features of 
point $B$ may be found by taking the limit $q^2/R^2 \rightarrow 0$ and then by 
putting the central pressure to infinity in the solutions. This furnishes 
$r_0/R = 2\sqrt{2}/3$. The limit for the mass function gives $m(0,r_0,R) = 
r_0^3/2R^2$ and then we get the  Buchdahl limit in the usual form, $r_0/m = 
9/4$.

\subsection{Point Q, $\boldsymbol{q=m=r_+=r_-=r_0=R}$: Quasiblack holes}

Point $Q$ represents quasiblack hole configurations, i.e., extremal charged 
stars that are on the verge of being a black hole~\cite{lemoszanchin2010}, see 
Figs.~\ref{regions-detail} and~\ref{regions-detail-closer}.

Point $Q$ is found by taking the limits $q/R \rightarrow 1$ and  $r_0/R 
\rightarrow 1$, and it gives $q=m=r_+=r_-=r_0=R$. Point Q belongs to the 
boundary of four very different regions: Region (a) with its undercharged 
regular stars, region (b) of the overcharged regular stars, region (c) of 
overcharged singular solutions, regions (d) and (e) of regular black holes. It 
is also the convergence of several curves or lines: The Buchdahl-Andr\'easson 
curve $C_4$, the dust curve $C_0$, the curve $C_1$ of null Guilfoyle parameter 
$a=0$, the second branch of the dust curve $C_2$, the curve $C_3$, the line 
$r_+=r_0 = R$, and the line of regular black holes with matching at the inner 
horizon $r_0=r_-=R$. This means that the solutions corresponding to this point 
are very special and highly degenerated, they are quasiblack holes, see also 
Sec.~\ref{r0=r+=r-} and \cite{lemoszanchin2010}.

\section{Conclusions}
\label{sec-conclusion}

We have explored the full parameter space of Guilfoyle's exact solutions for 
relativistic charged spheres \cite{guilfoyle}.

There are three free parameters among the various parameters of the model. We 
have chosen the normalized charge squared parameter $q^2/R^2$ and the 
normalized radius of the spheres $r_0/R$ as the good normalized parameters, 
where $R$ is the third free parameter characterizing the constant 
energy-density of the model, $8\pi\,\rho_m(r) + Q^2(r)/r^4 = 3/R^2$.  In order 
to avoid spacelike matching, the ratio $r_0/R$ is constrained in the interval 
$0\leq r_0/R \leq 1$. The limiting value $r_0/R =1$ implies the matching 
surface is lightlike, while $r_0/R<1$ leads to different kinds of spacetime.  
We allow the parameter $q^2/R^2$ to assume all values in the real line.  When 
$q^2$ is negative the interpretation of the parameter $q$ as electric charge 
is not possible.  However, $q^2$ may be interpreted as a tidal charge in a 
braneworld gravity.  The other two parameters, the mass $m$ and the Guilfoyle 
parameter $a$ depend on the three free parameters mentioned above.  The 
Guilfoyle parameter $a$ relates the metric to the gauge potentials.  Allowing 
$a$ to run along all values of the real line has proved essential in the 
search for new solutions.

The plethora of solutions found has been proved to bear very interesting 
spacetimes. Besides the well-behaved electrically charged stars examined in 
the original work \cite{guilfoyle}, for which it was shown that the fluid 
content has precise algebraic properties \cite{lemoszanchin2009}, there are 
quasiblack holes with pressure \cite{lemoszanchin2010}, there are charged 
stars that saturate the Buchdahl-Andr\'easson bound \cite{lemoszanchin2015} 
when the central pressure is allowed to reach arbitrarily large values, in the 
very same way the Schwarzschild interior solutions saturate the Buchdahl bound 
in the limit that the central pressure diverges, there are regular black holes 
with a de Sitter core and a coat of electric charge, regular black holes with 
a core of charged phantom fluid \cite{lemoszanchin2016}, other exotic regular 
black holes, as well as other solutions. This shows that the full spectrum 
presents a bewildering variety of possible solutions, from all types of stars 
to all types of black holes, though wormholes do not appear in the spectrum.
A stability analysis of Guilfoyle's solutions is yet to be performed.

\appendix
\section{Planar naked negative mass
singularity and Kasner metrics: Details}

\subsection{$\boldsymbol{r_0=r_+=0,\,  q^2< 0}$: Planar
naked negative mass singularities}
\label{1}

In the $r_0=r_+=0$ case, discussed in Sec.~\ref{r0=r+}, for a fixed
value of $q^2/R^2< 0$ one finds that the spacetime for small $r$ has
planar symmetry and is singular at $r=0$.  To see this, let us analyze
the limit $r_0/R \rightarrow 0$ in the Guilfoyle solutions for a fixed
value of $q^2/R^2< 0$.

Putting 
$r_0/R \rightarrow 0$ in the mass function, Eq.~\eqref{mass}, one finds
\begin{equation}
m= -\frac{|q|^2}{2r_0}+ {\rm O}\left(\frac{r_0^2}{|q|^2}\right)^{3/2}.
\label{mass-limit2}
\end{equation}
So, the mass is negative and divergent in this limit.

Putting $r_0/R \rightarrow 0$ in the equation for the Guilfoyle parameter 
$a$, Eq.~(\ref{a-function}), one finds
\begin{equation}
a= - \frac{|q|^2}{4r_0^2} - \frac{|q|^2}{4} + 
{\rm O}\left(\frac{r_0}{R}\right)^2\,.
\end{equation}
So, $a$ is negative and divergent in this limit.

Putting $r_0/R \rightarrow 0$ in the equation for the gravitational radius 
$r_+$, Eq.~(\ref{horizong}), using Eq.~(\ref{mass}), and $q^2=-|q|^2$, yields
\begin{equation}
r_+= r_0 -  {\rm O}\left(\frac{r_0^2}{|q|^2}\right)^{7/2}\,.
\label{rplus-limit2}
\end{equation}
Hence, in this limit the gravitational radius $r_+$ goes to zero as $r_0$, 
giving rise to a spacetime singularity at $r=r_0=r_+=0$.

Putting $r_0/R \rightarrow 0$ in the equation for the Cauchy radius $r_-$,
Eq.~(\ref{horizonc}), using Eq.~(\ref{mass}), and $q^2=-|q|^2$, yields 
\begin{equation}
r_-= -\frac{|q|^2}{2r_0} +r_0 + {\rm O}\left(\frac{r_0^2}{|q|^2}\right)^{3/2}\,.
\label{rminus-limit2}
\end{equation}
Hence, in this limit the Cauchy radius is negative and of no interest at all, 
since  we are considering solutions with $r\geq0$.

To interpret the resulting metric note that for $r_0$ very small but not zero, 
the metric for $r> r_0$ is the Reissner-Nordstr\"om metric with arbitrarily 
large and negative $m$, from Eq.~\eqref{mass-limit2}. This is equivalent to 
take the limit of large $-m$ with finite $q^2<0$ in the Reissner-Nordstr\"om 
metric, Eqs.~\eqref{metricsph}, \eqref{ABext1} and \eqref{ABext2}, yielding
\begin{equation}
 ds^2 = -\dfrac{2|m|}{r} dt^2 + \frac{r\, dr^2}{2|m|} +  r^2\,
(d\theta^2+\sin^2\theta\, d\varphi^2)\,.
\end{equation}
This metric is the negative mass Schwarzschild black hole metric close to
the singularity. It can be put into a form that makes explicit the
planar symmetry. Defining a new spacelike coordinate $l$ such that
$dl^2 = {r\, dr^2}/{2|m|}$ we get
$
 ds^2 =  -   \left({l}/{l_0}\right)^{\!\!\,\,-2/3}dt^2+ dl^2+
\left({l}/{l_0}\right)^{\!\!\,\,4/3}\left(4|m|^2\right)
(d\theta^2+\sin^2\theta\, d\varphi^2),
$
where we defined $l_0 = 4|m|/3$. To complete the transformation into a
plane-symmetric metric, note that at a given point with coordinates 
$(t,\, l,\, \theta_0,\, \varphi_0)$ we may define local Cartesian coordinates
$x=2m\left(\theta-\theta_0\right)$ and $y=2m\left(\varphi-\varphi_0\right)
\sin\theta_0$, where $(\theta,\, \varphi)$ are in the neighborhood of
$(\theta_0,\, \varphi_0)$, so the metric results in
 \begin{equation} \label{naked-plane}
 ds^2 = -   \left(\frac{l}{l_0}\right)^{\!\!-2/3}dt^2+ dl^2+
\left(\frac{l}{l_0}\right)^{\!\!4/3}\left(dx^2 + dy^2\right),
\end{equation}
which shows the planar symmetry of the surfaces defined by $t, \,l = {\rm 
constant}$. This metric can be considered a static Kasner metric (see, e.g., 
\cite{hiscock:1997jt,gpbook}), derived from the usual Kasner metric by a 
complex transformation that swaps the time coordinate with the radial one 
\cite{gpbook}, with characteristic Kasner parameters given by $p_1=-1/3$, and 
$p_2=p_3=2/3$. The radius of the matter distribution region $r_0$ can be 
finally set to zero, so that the two-dimensional timelike boundary surface 
$r=r_0= r_+$ flattens to a plane at $l=0$, giving rise to a timelike 
plane-symmetric singularity. The plane symmetric singularity at $l=0$ is 
naked.

\subsection{$\boldsymbol{r_0=r_-=0}$,  $q^2> 0$:
Kasner spacetimes}
\label{2}

In the $r_0=r_-=0$ case, discussed in Sec.~\ref{r0=r-}, for a fixed value of 
$q^2/R^2< 0$ one finds that the spacetime for small $r$ is a Kasner spacetime. 
 To see this, let us analyze the limit $r_0/R \rightarrow 0$ in the Guilfoyle 
solutions for a fixed value of $q^2/R^2< 0$.

Putting 
$r_0/R \rightarrow 0$ in the mass function, Eq.~\eqref{mass}, one finds
\begin{equation}
m= \frac{q^2}{2r_0}+ {\rm O}\left(\frac{r_0^2}{q^2}\right)^{3/2}\,.
\label{mass-limit}
\end{equation}
So, for finite charge and in the limit $r_0/R\rightarrow 0$, the mass
inside $r_0$ is positive and diverges giving rise to a spacetime singularity 
at $r_-=r_0=0$.

Putting 
$r_0/R \rightarrow 0$ in the equation for the Guilfoyle parameter $a$, 
Eq.~(\ref{a-function}), one finds
\begin{equation}
a=  \frac{q^2}{4r_0^2} +\frac{q^2}{4} + 
{\rm O}\left(\frac{r_0}{R}\right)^2\,.
\end{equation}
So, $a$ is positive and divergent in this limit.

Putting 
$r_0/R \rightarrow 0$ in the equation for the gravitational radius $r_+$,
Eq.~(\ref{horizong}), using Eq.~(\ref{mass}) for $q^2>0$, yields 
\begin{equation}
r_+= \frac{q^2}{2r_0} -r_0 + {\rm O}\left(\frac{r_0^2}{q^2}\right)^{3/2}\,.
\label{rplus-limit}
\end{equation}
Hence, in this limit the gravitational radius $r_+$ diverges.

Putting $r_0/R \rightarrow 0$ in the equation for the Cauchy radius $r_-$,
Eq.~(\ref{horizonc}), using Eq.~(\ref{mass}) for $q^2$, yields 
\begin{equation}
r_-= r_0 +  {\rm O}\left(\frac{r_0^2}{q^2}\right)^{7/2}\,.
\label{rminus-limit}
\end{equation}
Hence, in this limit the Cauchy radius goes to zero as $r_0$, it disappears.

To interpret the resulting metric note that for $r_0$ very small but
not zero, the metric for $r> r_0$ is the Reissner-Nordstr\"om metric
with arbitrarily large and positive $m$, from Eq.~\eqref{mass-limit2}.
This is equivalent to take the
limit of large $m$ with finite $q^2>0$ in the Reissner-Nordstr\"om metric,
Eqs.~\eqref{metricsph}, \eqref{ABext1} and \eqref{ABext2}, yielding
\begin{equation}
ds^2 = +\dfrac{2m}{r} dt^2 - \frac{r\, dr^2}{2m} + r^2(d\theta^2
+\sin^2\theta\, d\varphi^2)\,.
\label{k1}
\end{equation}
This metric is the Schwarzschild black hole metric close to the singularity. 
It can be put into a form that is locally a Kasner metric (see, e.g., 
\cite{hiscock:1997jt}). Defining a new timelike coordinate $\tau$ such that 
$d\tau^2 = {r\, dr^2}/{2m}$, and a new spacelike coordinate $r$ by $r=t$ we 
get
$
ds^2 = -d\tau^2 +  \left({\tau}/{\tau_0}\right)^{\!\!\,\,-2/3}dr^2+
\left({\tau}/{\tau_0}\right)^{\!\!\,\,4/3}\left(4m^2\right)(d\theta^2
+\sin^2\theta\, d\varphi^2)
$,
where we defined $\tau_0 = 4m/3$. To complete the transformation into a
plane-symmetric metric, note that at a given point with coordinates 
$(t,\, l,\, \theta_0,\, \varphi_0)$ we may define local Cartesian coordinates
$x=2m\left(\theta-\theta_0\right)$ and $y=2m\left(\varphi-\varphi_0\right)
\sin\theta_0$, where $(\theta,\, \varphi)$ are in the neighborhood of
$(\theta_0,\, \varphi_0)$, so the metric results in
\begin{equation}
ds^2 = -d\tau^2 +  \left(\frac{\tau}{\tau_0}\right)^{-2/3}dr^2+
\left(\frac{\tau}{\tau_0}\right)^{4/3}\left(dx^2 + dy^2\right),
\end{equation}
which is the Kasner universe metric with characteristic Kasner 
parameters given by $p_1=-1/3$, and $p_2=p_3=2/3$
(see, e.g., \cite{hiscock:1997jt,gpbook}).
The radius of the matter distribution $r_0$ can be finally set to zero, so 
that the two-dimensional timelike boundary surface at $r=r_0$ flattens to a 
plane containing an infinite mass, giving rise to a spacelike singularity. The 
singularity is at $\tau=0$, a big-bang or a big-crunch like singularity. The 
surfaces $\tau$, $r = {\rm constant}$ are now two-dimensional spacelike 
planes.

\section*{Acknowledgments}

We thank Funda\c c\~ao para a Ci\^encia e Tecnologia (FCT), Portugal,
for financial support through Grant No.~UID/FIS/00099/2013.  This work
was partially funded by Coordena\c{c}\~ao de Aperfei\c{c}oamen\-to do
Pessoal de N\'\i vel Superior (CAPES), Brazil, Grants
No.~88881.064999/2014-01 and Grant No.~88887.068694/2014-00.
J. P. S. L. also thanks an FCT grant,
No.~SFRH/BSAB/128455/2017, 
and Piotr Chru\'sciel and the 
Gravitational Physics Group at the 
Faculty of Physics, University of Vienna,
for hospitality.
V. T. Z. 
thanks Funda\c c\~ao de Amparo \`a Pesquisa do Estado de S\~ao Paulo (FAPESP),
Grant No.~2011/18729-1, and Conselho Nacional de
Desenvolvimento Cient\'\i fico e Tecnol\'ogico of Brazil (CNPq), Grant
No.~308346/2015-7.

\end{document}